\pdfoutput=1
\documentclass[letterpaper,preprints,article,accept,moreauthors,pdftex]{Definitions/mdpi}

\firstpage{1}
\pubvolume{1}
\issuenum{1}
\articlenumber{0}
\pubyear{2021}
\copyrightyear{2021}
\datereceived{}
\dateaccepted{}
\datepublished{}
\hreflink{https://doi.org/10.3390/a14030090} 

\usepackage[ruled, vlined, linesnumbered]{algorithm2e}
\usepackage{makecell}
\usepackage{subcaption}
\usepackage{tabulary}
\usepackage{rotating}

\usetikzlibrary{positioning,calc,shapes,matrix}
\tikzstyle{snode}=[circle,inner sep=0.5mm, minimum size=2.5mm,draw = black, fill=black]
\tikzstyle{node}=[circle,inner sep=0.5mm,minimum size=5.25mm,draw = black]

\newcommand{\tdcch}{CATCHUp}

\Title{Space-Efficient, Fast and Exact Routing in Time-Dependent Road Networks $^{\dagger}$}
\TitleCitation{Space-Efficient, Fast and Exact Routing in Time-Dependent Road Networks}

\Author{Ben Strasser, Dorothea Wagner and Tim Zeitz*\orcidA{}}

\AuthorNames{Ben Strasser, Dorothea Wagner and Tim Zeitz}
\AuthorCitation{Strasser, B.; Wagner, D.; Zeitz, T.}

\address{%
Institute of Theoretical Informatics, Karlsruhe Institute of Technology, Am Fasanengarten 5,76131 Karlsruhe, Germany; academia@ben-strasser.net (B.S.); dorothea.wagner@kit.edu (D.W.); tim.zeitz@kit.edu (T.Z.)}

\corres{Correspondence: tim.zeitz@kit.edu;
}

\conference{Proceedings of the 28th Annual European Symposium on Algorithms (ESA 2020)} 

\abstract{
We study the problem of quickly computing point-to-point shortest paths in massive road networks with traffic predictions.
Incorporating traffic predictions into routing allows, for example, to avoid commuter traffic congestions.
Existing techniques follow a two-phase approach:
In a preprocessing step, an index is built.
The index depends on the road network and the traffic patterns but not on the path start and end.
The latter are the input of the query phase, in which shortest paths are computed.
All existing techniques have large index size, slow query running times or may compute suboptimal paths.
In this work, we introduce \tdcch{} (Customizable Approximated Time-dependent Contraction Hierarchies through Unpacking), the first algorithm that simultaneously achieves all three objectives.
The core idea of \tdcch{} is to store paths instead of travel times at shortcuts.
Shortcut travel times are derived lazily from the stored paths.
We perform an experimental study on a set of real world instances and compare our approach with state-of-the-art techniques.
Our approach achieves the fastest preprocessing, competitive query running times and up to 38 times smaller indexes than competing approaches.
}

\keyword{realistic road networks; time-dependent route planning; shortest paths}  

\begin{document}

\section{Introduction}

\begin{figure}
\centering
\begin{tikzpicture}[]
  \node [snode] at (0, 0) (v1) {};
  \node [snode] at (1.5, 0.8) (v2) {};
  \node [snode] at (3.2, -0.5) (v3) {};
  \node [snode] at (3, 1) (v4) {};
  \node [snode] at (5, 1.2) (v5) {};
  \node [snode] at (5.3, 0.2) (v6) {};

  \node [] at (10, .2) (ttf) {\includegraphics[width=180pt]{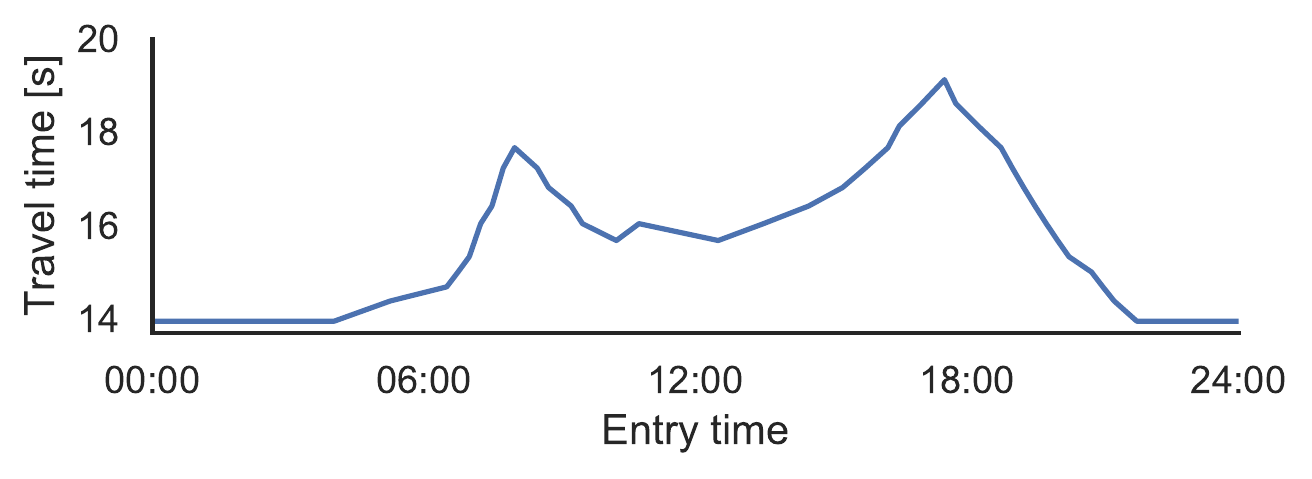}};

  \draw [->] (v1) -- (v2);
  \node at (0.3, 0.5) {\includegraphics[width=6mm]{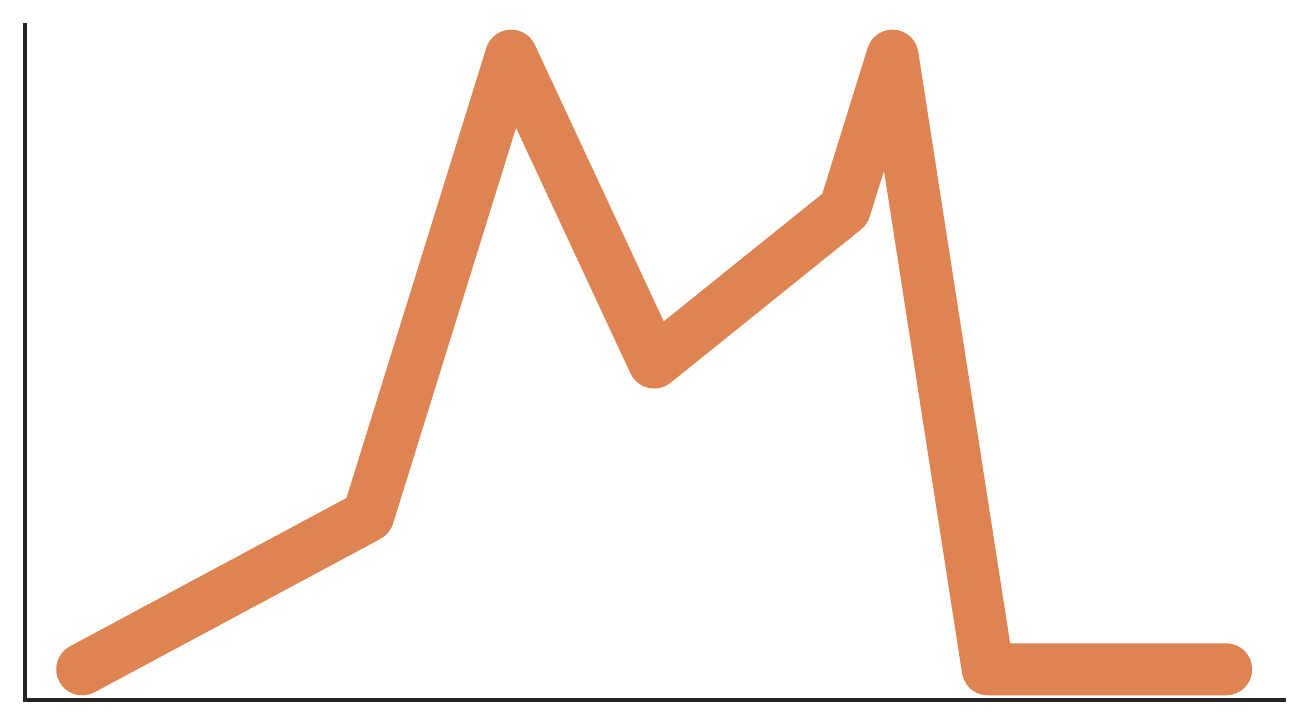}};
  \draw [->] (v2) -- (v3);
  \node at (2.5, 0.5) {\includegraphics[width=6mm]{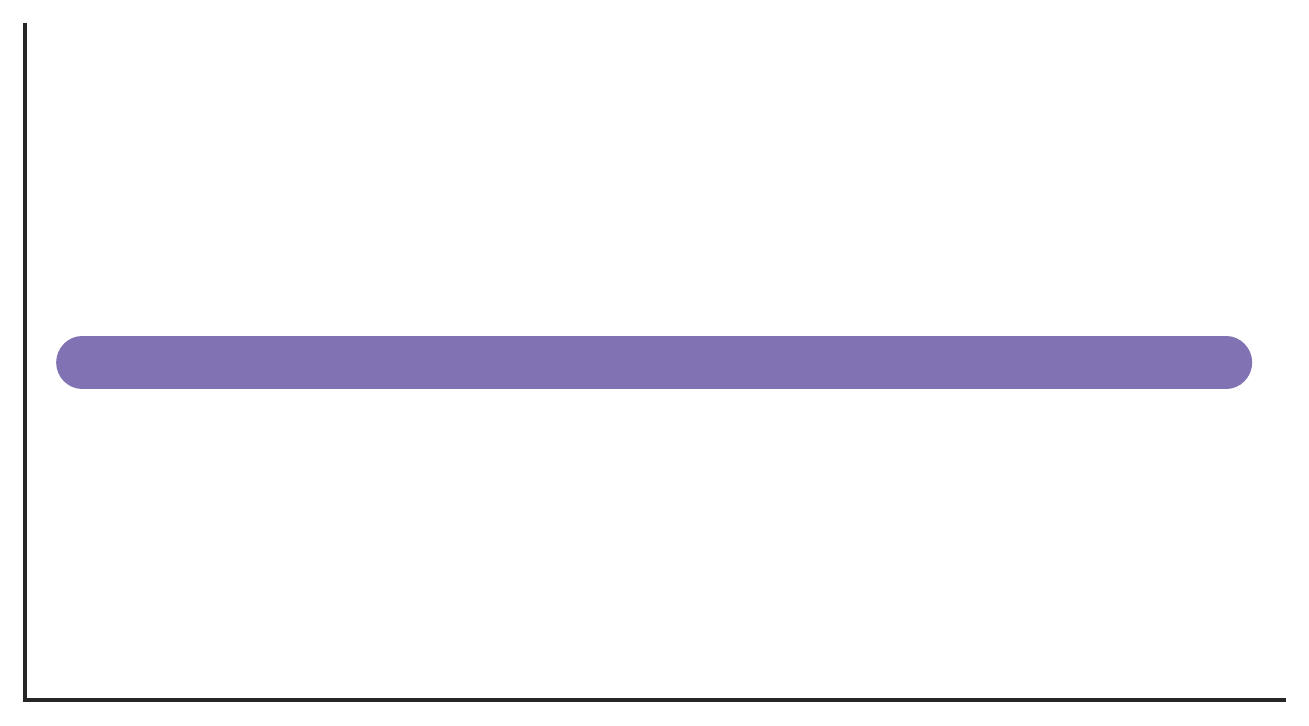}};
  \draw [->] (v1) -- (v3) node[midway,below] {\includegraphics[width=6mm]{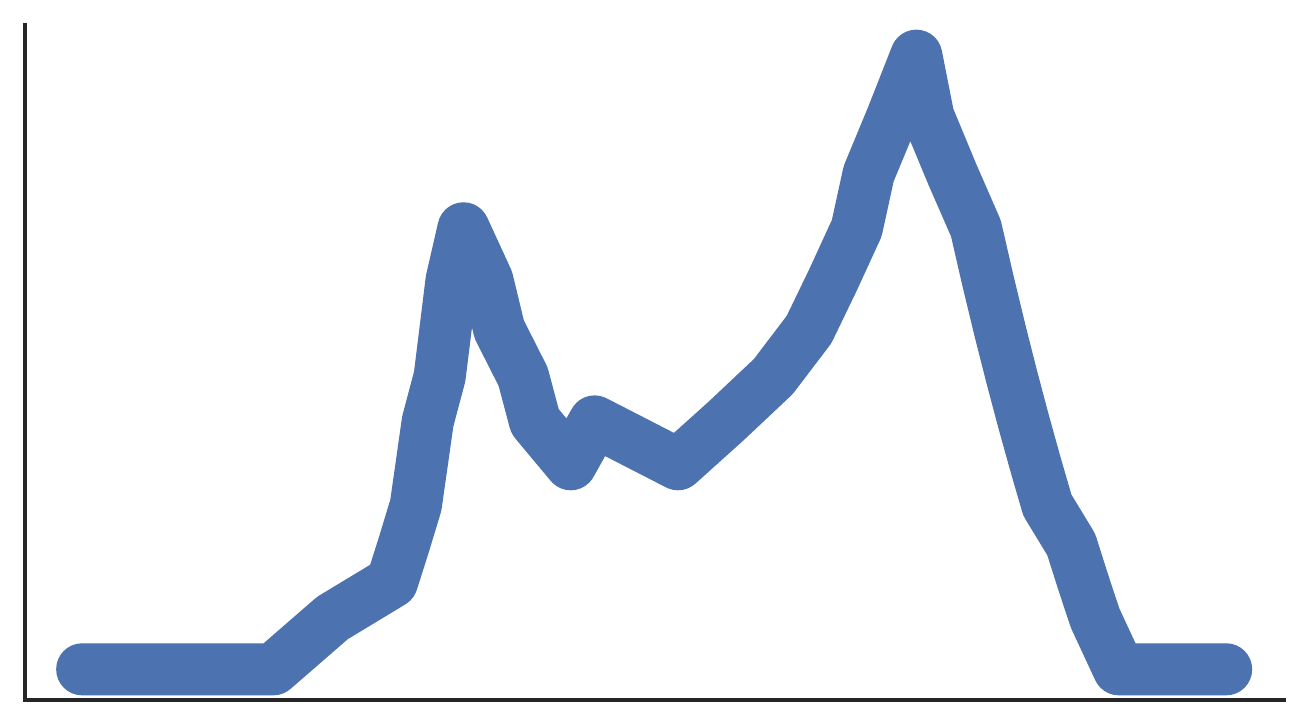}};
  \draw [<->] (v3) -- (v4) node[midway,right] {\includegraphics[width=6mm]{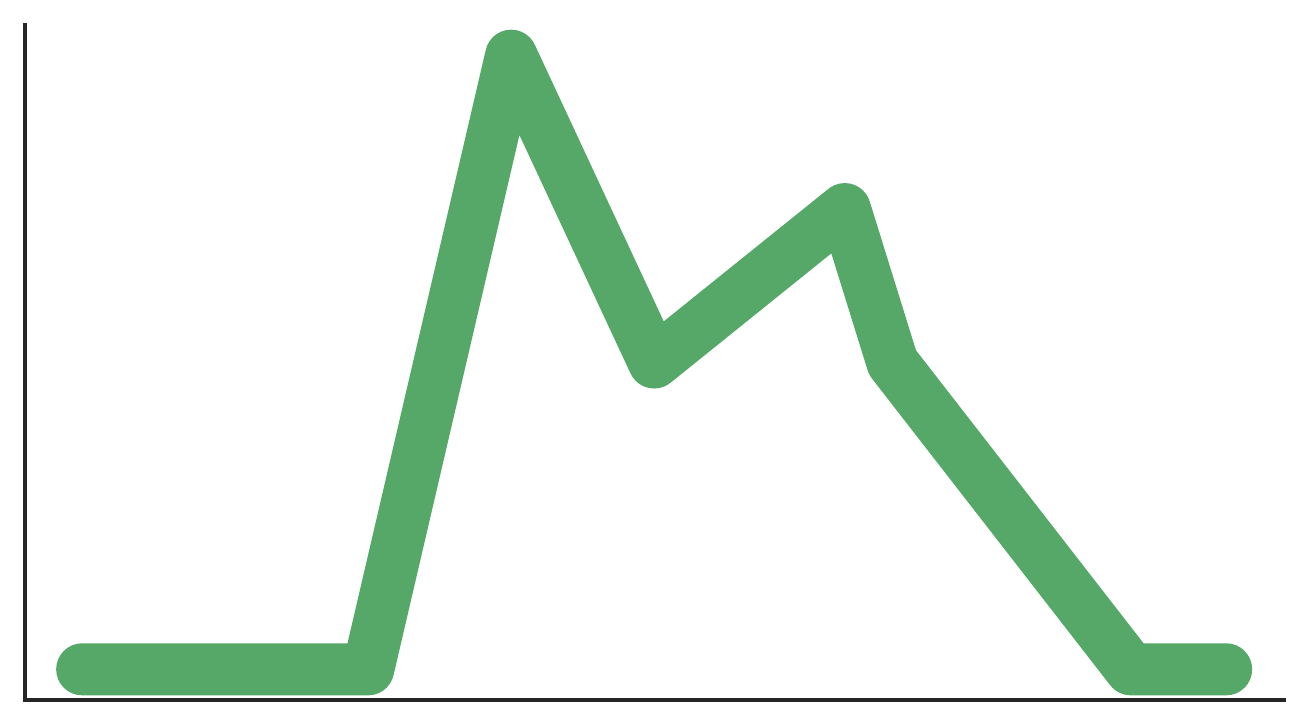}};
  \draw [<->] (v2) -- (v4) node[midway,above] {\includegraphics[width=6mm]{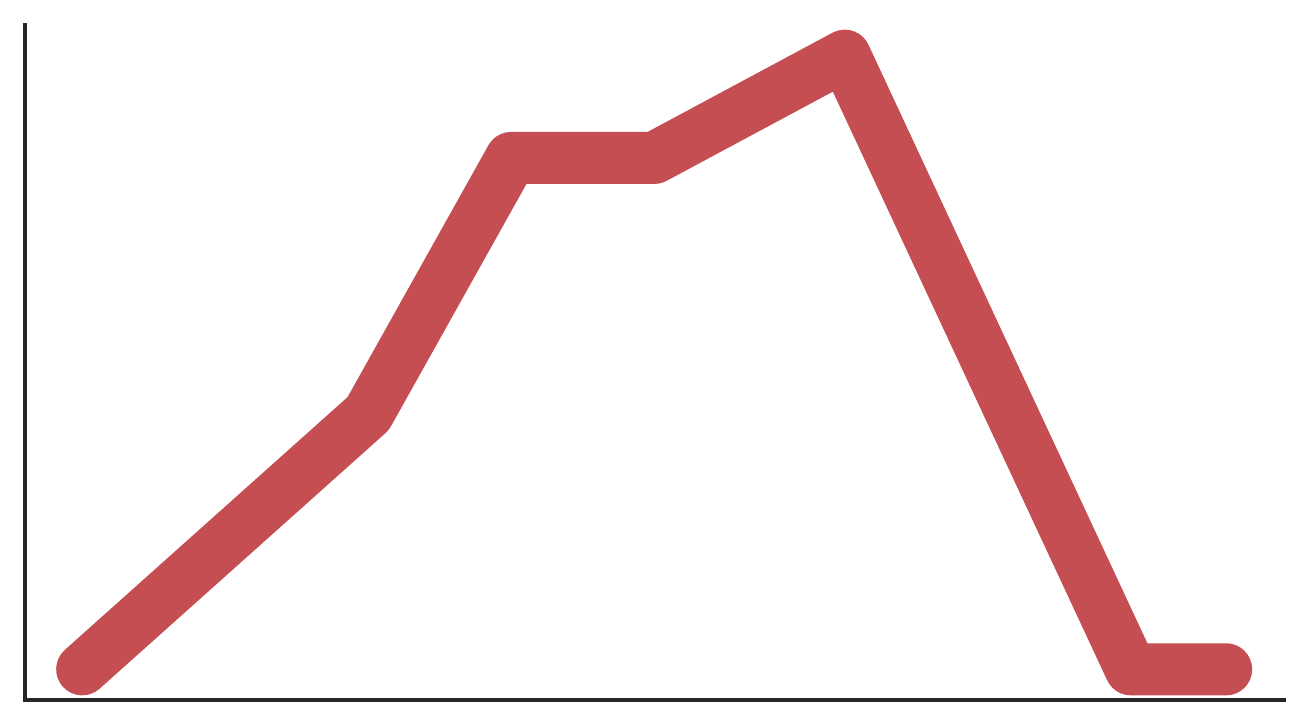}};
  \draw [->] (v4) -- (v5) node[midway,above] {\includegraphics[width=6mm]{fig/ttf4_stripped.pdf}};
  \draw [->] (v5) -- (v6);
  \node at (5.6, .7) {\includegraphics[width=6mm]{fig/ttf_stripped.pdf}};
  \draw [<->] (v3) -- (v6) node[midway,below] {\includegraphics[width=6mm]{fig/ttf2_stripped.pdf}};

  \draw [-, dashed] (5.95, .87) -- (7.4, 1.5);
  \draw [-, dashed] (5.95, .53) -- (7.4, -.7);
\end{tikzpicture}
\caption{The graph of a small road network with predicted travel times for each road segment.}\label{fig:tdgraph}
\end{figure}

Routing in road networks is a well-studied topic with a plethora of real world applications.
Services such as Google, Baidu, Yandex, Bing, Apple, or HERE Maps are ubiquitous and used by millions of users on a daily basis.
The core problem is to compute the fastest route between a source and a target.
The idealized problem can be formalized as the classic point-to-point shortest path problem.
Streets are modeled as arcs.
Street intersections are modeled as nodes.
Travel times are modeled as scalar arc weights.
Unfortunately, this idealized view does not model certain important real world effects.
An example are recurring commuter congestions.
In this article, we consider an extended problem in which travel times are time-dependent.
The travel time of an arc is a function of the moment where a car enters the arc.
Figure~\ref{fig:tdgraph} depicts an example.

Computing shortest-paths using Dijkstra's \cite{d-ntpcg-59} algorithm is possible both in the classical and in the time-dependent setting.
However, for many applications, its running time is too large.
To achieve fast running times, a two-phase approach is used.
In the first phase, the \emph{preprocessing} phase, an index is constructed.
The index only depends on the road networks and the arc travel times.
In the second phase, the \emph{query} phase, shortest paths are computed utilizing this index.

\begin{figure}
\centering
\begin{subfigure}[b]{0.48\columnwidth}
\begin{tikzpicture}[]
  \node [snode] at (1, 0) (v1) {};
  \node [snode] at (2, 0) (v2) {};
  \node [snode] at (3, 0) (v4) {};
  \node [snode] at (4, 0) (v5) {};
  \node [snode] at (5, 0) (v6) {};
  \node [snode] at (3.5, -0.5) (v7) {};
  \node [snode] at (4.2, -0.5) (v8) {};

  \draw [->] (.2,.8) -- (v1);
  \draw [->] (0,0) -- (v1);
  \draw [->] (.2,-.8) -- (v1);

  \draw [->] (v1) -- (v2);
  \draw [->] (v2) -- (v4);
  \draw [->] (v4) -- (v5);
  \draw [->] (v5) -- (v6);

  \draw [->] (v2) -- (v7);
  \draw [->] (v7) -- (v8);
  \draw [->] (v8) -- (v6);

  \draw [->] (v6) -- (5.8, .5);
  \draw [->] (v6) -- (5.8, -.5);

  \draw [->, dashed] (v1) to [bend left=35] (v6);
  \draw [->, dashed, draw=black!50] (v2) to [bend left=35] (v6);
  \draw [->, dashed, draw=black!40] (v2) to [bend left=35] (v5);
  \draw [->, dashed, draw=black!40] (v7) to (v6);
\end{tikzpicture}
\caption{A shortcut arc (dashed, black) bypassing several nodes. In this work, shortcuts always skip over exactly one node and two arcs, which may in turn be shortcut arcs (dashed gray arcs).}\label{fig:shortcut}
\end{subfigure}
\hspace{1mm}
\begin{subfigure}[b]{0.48\columnwidth}
\includegraphics[width=\columnwidth]{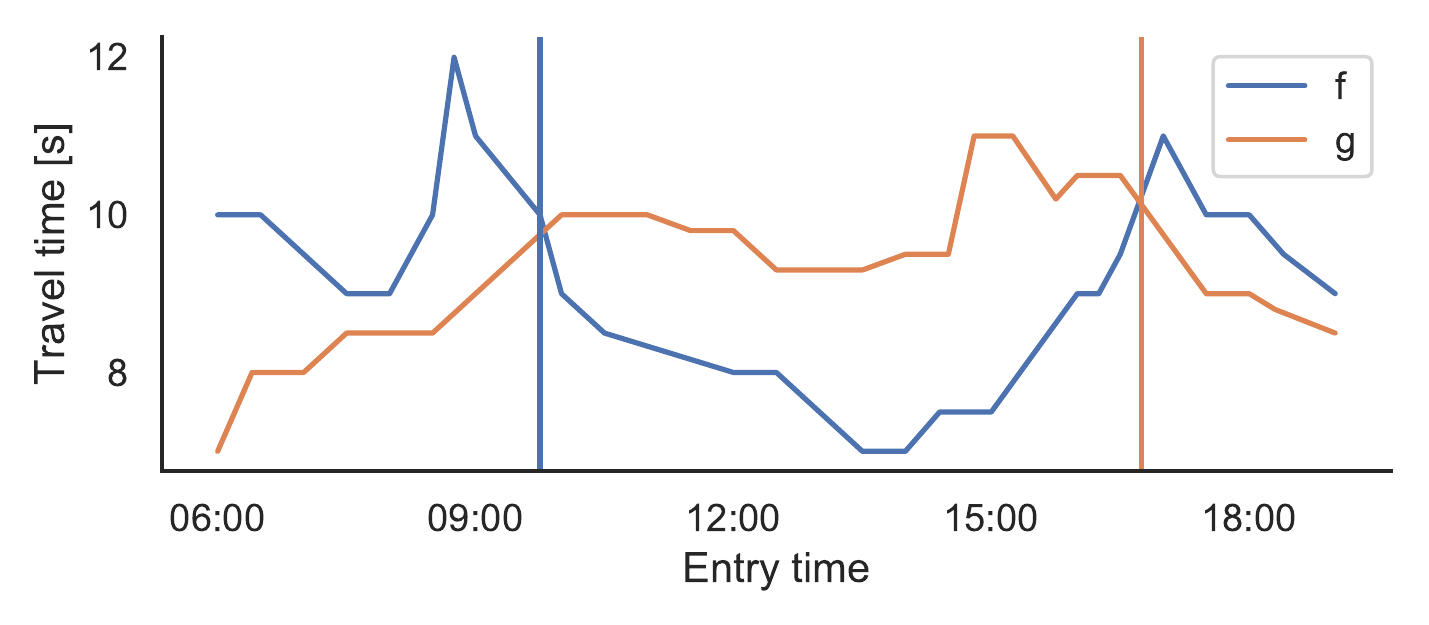}
\caption{Travel time functions for two different paths between the same start and end node.}\label{fig:compression}
\end{subfigure}
\caption{Shortcuts and their travel time functions.}
\end{figure}
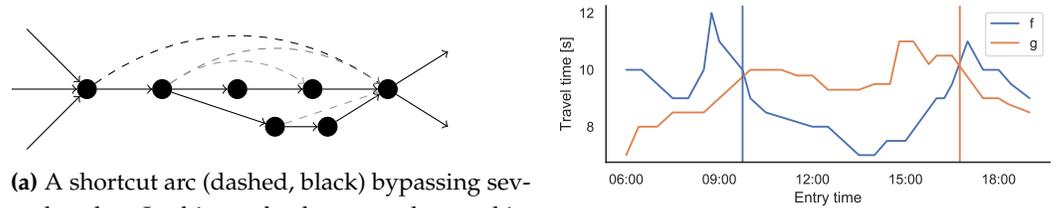

An important ingredient for many two-phase techniques \cite{bdgmpsww-rptn-16,bd-sharc-09,dn-crdtd-12,dsw-cch-15,gssv-erlrn-12} are \emph{shortcuts}.
Shortcuts are additional arcs introduced during preprocessing, which bypass parts of the input graph such as in Figure~\ref{fig:shortcut}.
The weight of a shortcut is set to the length of the shortest path between its endpoints.
When computing shortest paths, only few shortcuts are explored instead of many arcs in the input graph.
The path represented by a shortcut can be obtained lazily, for example by running local Dijkstra searches~\cite{dgpw-crprn-13}, or by iterating over possible middle nodes when shortcuts always represent two other (shortcut) arcs~\cite{dsw-cch-15,gssv-erlrn-12}.

This approach has been extended to the time-dependent setting~\cite{bgsv-mtdtt-13,bdpw-dtdrp-16}.
Shortcuts are no longer associated with scalar weights.
Instead, \emph{travel time functions} are used that map the entry time into a shortcut to the travel time through it.
Typically, these functions are represented as piecewise linear functions.
They are stored as a sequence of \emph{breakpoints}.
Unfortunately, these functions can become very complex.
Computing and storing them is expensive.
The number of breakpoints in a shortcut's function practically corresponds to the accumulated number of breakpoints of the functions of bypassed arcs.
Contrary to the classic setting, shortcuts aggregate the complexity of paths they represent, rather than skipping it.
This leads to slow preprocessing and prohibitive memory consumption.

In this paper, we explore an alternative approach to shortcut travel time functions.
Rather than explicitly storing them and obtaining paths lazily, we store paths and obtain travel times lazily.
We expect that the shortest path between two nodes changes less frequently than the travel time.
Intuitively, going via a highway may be slower due to congestion but is usually still the fastest option.
Consider the functions $f$ and $g$ in Figure~\ref{fig:compression}.
These functions are travel time functions of two paths between the same endpoints and have many breakpoints. 
If we want to store the travel time function of a shortcut between these endpoints, we need to store the function $h = \min(f, g)$.
Storing $h$ explicitly requires roughly a number of breakpoints proportional to the number of breakpoints in $f$ and $g$.
However, if we only store which path is the fastest, we only need to store the points in time when the faster path switches.
We expect significantly fewer switches than breakpoints.
In this paper, we employ this alternative approach to adapt an existing speed-up technique to the time-dependent setting, describe engineering techniques employed in our implementation, and present experimental results demonstrating that our approach significantly reduces memory consumption while achieving competitive query times.

\subparagraph*{Related Work}

Route planning in road networks has been extensively studied in the past decade.
An overview over the field can be found in \cite{bdgmpsww-rptn-16}.
Here, we focus on speed-up techniques for time-dependent road networks.
Several time-independent speed-up techniques have been generalized to the time-dependent setting.
ALT \cite{gh-cspas-05}, an approach using landmarks to obtain good A* \cite{hnr-afbhd-68} potentials has been generalized to TD-ALT \cite{ndls-bastd-12} and successively extended with node contraction to TD-CALT \cite{dn-crdtd-12}.
Even when combined with approximation, TD-CALT queries may take longer than 10\,ms on continental sized graphs.
SHARC \cite{bd-sharc-09}, a combination of ARC-Flags \cite{l-aefea-04} with shortcuts which allows unidirectional queries was also extended to the time-dependent scenario \cite{d-tdsr-11}.
It can additionally be combined with ALT yielding L-SHARC \cite{d-tdsr-11}.
SHARC can find short paths in less than a millisecond but does not always find a shortest path.
MLD/CRP \cite{dgpw-crprn-13,hsw-emlog-08} has been extended to TD-CRP \cite{bdpw-dtdrp-16} which can be used in a time-dependent setting. 
TD-CRP requires approximation to achieve reasonable memory consumption.
It may find suboptimal paths.
Another approach is FLAT \cite{kmppwz-eotdr-16} and its extension CFLAT \cite{kppwz-iotdr-17a}.
CFLAT features sublinear query running time after subquadratic preprocessing and guarantees on the approximation error.
Similar to our approach, CFLAT uses timestamped combinatoric structures to represent the changes in shortest paths over time and computes travel times lazily.
Unfortunately, preprocessing takes long in practice and generates a prohibitively large index size.

There are several approaches based on Contraction Hierarchies~\cite{gssv-erlrn-12}.
Three were introduced by \citet{bgsv-mtdtt-13}: Time-dependent CH (TCH), inexact TCH, and Approximated TCH (ATCH).
TCH achieve great query performance but at the cost of a huge index size on state-of-the-art continental sized instances.
The index size can be reduced at the cost of exactness (inexact TCH) or query performance (ATCH).
An open-source reimplementation of~\cite{bgsv-mtdtt-13} named KaTCH\footnote{\url{https://github.com/GVeitBatz/KaTCH}} exists.
A simple heuristic named Time-Dependent Sampling (TD-S) was introduced by \citet{s-dtdrr-17}.
It samples a fixed set of scalar values from the time-dependent functions.
It has manageable index sizes and fast query times but does not always find shortest paths.

Time-dependent shortest path algorithms are sometimes used as a subroutine in time-dependent vehicle routing problems~(VRP)~\cite{huang2017time}.
VRP is a large field and discussing it beyond the scope of this article.
For an overview over time-dependent VRP and other time-dependent routing problem variants we refer to the work by \citet{gendreau2015time}.

\subparagraph*{Contribution}

In this work, we explore a variant of time-dependent Contraction Hierarchies, where shortcuts store paths instead of travel times.
We introduce \tdcch{} (Customizable Approximated Time-dependent Contraction Hierarchies through Unpacking), a time-dependent generalization of Customizable Contraction Hierarchies~\cite{dsw-cch-15} and a thoroughly engineered implementation.
Preprocessing takes only minutes even on modern production-grade continental sized instances with tens of millions of nodes.
We also present algorithms which allow us to employ approximation to accelerate preprocessing without sacrificing exactness for the queries.
Our implementation achieves fast and exact queries with performance competitive to TCH queries while requiring up to 38 times less memory.

This paper is an extended version of a conference paper~\cite{swz-sfert-20}.
In addition to the previously reported results, we describe our algorithms in greater depth and provide additional important engineering details.
We also introduce new algorithms for profile queries.
The experimental evaluation has been significantly extended.
We perform experiments with more graphs and provide a deeper analysis on the performance of our preprocessing algorithms.

\section{Materials and Methods}

In this section, we describe our algorithms, data structures and implementation.
After introducing preliminaries, we describe existing algorithms we build upon, most importantly Customizable Contraction Hierarchies.
Section~\ref{sec:shortcut} introduces our shortcut unpacking data structure which we use to efficiently reconstruct the paths represented by a shortcut.
We continue by presenting our our algorithms.
The preprocessing, which computes auxiliary data from a road network with traffic predictions, is discussed in Section~\ref{sec:preprocessing}.
In Section~\ref{sec:queries}, we present query algorithms which utilize the auxiliary data to efficiently compute shortest travel times and paths between two given location.

\subsection{Preliminaries}\label{sec:prelim}
We model road networks as directed graphs $G=(V,A)$.
A node $v \in V$ represents an intersection and an arc $a=uv \in A$ with $u,v \in V$ represents a road segment.
Every arc $a$ has a \emph{travel time function} $f_a: \mathbb{R} \to \mathbb{R}^{>0}$ mapping departure time to travel time.
These functions are also referred to as \emph{travel time profiles}.
We assume that travel time functions fulfill the \emph{First-In-First-Out} (FIFO) property, that is, for any $\sigma, \tau \in \mathbb{R}$ with $\sigma \leq \tau$, $\sigma + f(\sigma) \leq \tau + f(\tau)$ has to hold.
Informally, this means that it is not possible to arrive earlier by starting later.
If there are arcs that do not fulfill the FIFO property, the shortest path problem becomes $\mathcal{NP}$-hard~\cite{or-tnp-89} if waiting is not allowed.
In our implementation, travel time functions are periodic piecewise linear functions represented by a sequence of \emph{breakpoints}.
We denote the number of breakpoints in a function $f$ by its \emph{complexity} $|f|$.
A path is a sequence of nodes $[v_1, \dots, v_k]$ such that $v_i v_{i+1} \in A$.
We denote the concatenation of two paths by $[v_1, \dots, v_k] \cdot [v_k, \dots, v_l] = [v_1, \dots, v_k, \dots v_l]$.
The travel time to traverse a path $[v_1, \dots, v_k]$ can be evaluated by successively evaluating each link's travel time:
$\textsc{Eval}([v_1, \dots, v_k], \tau) = f_{v_1 v_2}(\tau) + \textsc{Eval}([v_2, \dots, v_k], \tau + f_{v_1 v_2}(\tau))$.

Given two travel time functions $f$ and $g$ for arcs $uv$ and $vw$, we are often interested in the travel time function of traversing first $uv$ and then $vw$, that is $f(\tau) + g(f(\tau) + \tau)$.
Computing this function is called \emph{linking}.
In a slight abuse of notation, we write $g \circ f$ for this linked function.
When combining two travel time functions $f$ and $g$ for different paths $[u,\dots,v]$ with the same start and end, we often want to know the travel time of the best path between $u$ and $v$, that is $\min(f, g)$.
Computing this function is called \emph{merging}.
Both linking and merging can be implemented with coordinated linear sweeps over the breakpoints of both functions. 

Given a departure time $\tau$ and nodes $s$ and $t$, an \emph{earliest-arrival query} asks for earliest point in time one can arrive at $t$ when starting from $s$ at $\tau$.
Such a query can be handled by Dijkstra's algorithm~\cite{d-ntpcg-59} with minor modifications~\cite{d-aassp-69}.
The algorithm keeps track of the earliest known arrival time $\operatorname{ea}_v$ at each node $v$.
These labels are initialized with $\tau$ for $s$ and $\infty$ for all other nodes.
A priority queue is initialized with $(s,\tau)$.
In each step, the node $u$ with minimal earliest arrival $\operatorname{ea}_u$ is popped from the queue and outgoing arcs are \emph{relaxed}.
To relax an arc $uv$, the algorithm checks if $\operatorname{ea}_u + f_{uv}(\operatorname{ea}_u)$ improves $\operatorname{ea}_v$ and updates label and queue position of $v$ accordingly.

When nodes are popped from the queue, their earliest arrival is final.
This property is denoted as \emph{label-setting}.
Once $t$ is extracted from the queue, the earliest arrival at $t$ is known.
To retrieve the shortest path, one can use \emph{parent pointers} which store the previous node on the shortest path from $s$ for each node.
We refer to this algorithm as \emph{TD-Dijkstra}.

A \emph{profile query} asks for the shortest travel time function between two nodes $s$ and $t$ for a time interval $T$.
This query type can also be solved by a variant of Dijkstra's algorithm.
For each node $v$, instead of the earliest arrival time a tentative travel time function $f_{v}$ from $s$ to $v$ is maintained.
Initially, $f_{v}(\tau)$ is set to $\infty$ for all $\tau \in T$ and to zero for $s$.
In the priority queue, nodes are ordered by the current lower bound of their travel time function.
The queue is initialized with $(s,0)$.
When a node is popped from the queue, outgoing arcs are relaxed.
Here, relaxing an arc $uv$ means linking $f_u$ with $f_{uv}$ and merging the result with the travel time function of $v$: $f_v = \min(f_v, f_{uv} \circ f_u)$.
If the travel time to $v$ can be improved, its label and queue position will be updated accordingly.

This algorithm is not label-setting.
Nodes may be popped several times from the queue.
The algorithm can terminate as soon as the minimum key in the queue is greater than $\max_{\tau \in T} f_t(\tau)$.
We refer to this algorithm as \emph{TD-Profile-Dijkstra}.

The \emph{A* algorithm} \cite{hnr-afbhd-68} is an extension to Dijkstra's algorithm.
It reduces the number of explored nodes by guiding the search towards $t$.
Each node $u$ has a potential $\rho_t(u)$ which is an estimate of the distance to $t$.
The priority queue is then ordered by $\operatorname{ea}_u + \rho_t(u)$.

\subsubsection{Contraction Hierarchies}

\emph{Contraction Hierarchies}~(CH)~\cite{gssv-erlrn-12} is a speed-up technique exploiting the inherent hierarchy in road networks.
It was initially developed for networks with scalar edge weights.
Nodes are heuristically ranked by their importance.
Nodes with higher rank should cover more shortest paths.
During preprocessing, all nodes are \emph{contracted} in order of ascending importance.
Contracting a node $v$ means removing it from the network but preserving all shortest distances among remaining higher ranked nodes.
For this, \emph{shortcut arcs} are inserted between the neighbors of $v$ if a shortest path goes through $v$.
A shortcut is only necessary if it represents the only remaining shortest path between its endpoints.
This can be checked with a local Dijkstra search (called \emph{witness search}) between the endpoints.
The result of the preprocessing is called an \emph{augmented graph}.

Queries can be answered by performing a bidirectional Dijkstra search on the augmented graph.
The forward search starts at $s$ and relaxes only forward arcs to higher ranked nodes.
The backward search starts at $t$ and traverses arcs in reverse direction and also only searches to higher ranked nodes.
The construction of the augmented graph guarantees that the searches will meet and find a path that has the same length as shortest paths in the original graph.
The higher ranked nodes reachable from a node are referred to as the node's \emph{CH search space}.

\subsubsection{Customizable Contraction Hierarchies}

\emph{Customizable Contraction Hierarchies}~(CCH)~\cite{dsw-cch-15} is a CH extension.
It splits CH preprocessing into two phases where only the second uses weights.
In the first phase, a separator decomposition and an associated nested dissection order~\cite{bcrw-s-16,g-ndrfe-73} are computed.
This order determines the node ranks.
Nodes in the top-level separators have the highest ranks, followed by the nodes of each cell, recursively ordered by the same method.
Since all shortest paths between different cells have to use separator nodes, these nodes cover many shortest paths.
Thus, a nested dissection order is a good CH order.

Then, nodes are contracted iteratively ordered by ascending rank.
Because weights are not considered in this phase, no witness search can be performed.
All potential shortcuts between the higher ranked neighbors of the current node will be inserted.
The upward neighbors become a clique.
This phase is performed on the bidirected input graph with arcs $A' = A \cup \{ vu : uv \in A \}$ where each arc exists in both directions.

In the second phase called \emph{customization}, arc weights are computed.
Arcs in the augmented graph corresponding to an input graph arc are initialized with the corresponding weight.
All other arc weights are set to $\infty$.
Then, all arcs are processed in ascending order of their lower ranked endpoint.
To process an arc $uv$, all \emph{lower triangles} $[u,w,v]$, where $w$ has lower rank than $u$ and $v$ are enumerated, checking if the path $[u,w,v]$ is shorter than $uv$.
If so, the weight of $uv$ is set to the length of the $[u,w,v]$.
We denote this as \emph{lower triangle relaxation}.

The result of this basic customization fulfills all necessary properties for the CH query algorithm to find correct shortest distances.
However, it contains many unnecessary arcs.
These can optionally be removed using the \emph{perfect customization} algorithm.
Here, all arcs are processed in descending order of their higher ranked endpoint.
For each arc $uv$, upper and intermediate triangles are enumerated, i.e., the third node $w$ has greater rank than either $u$ or $v$.
The weight of $uv$ will then be decreased if possible to the weight of the path $[u,w,v]$.
Once all arcs have been processed, all arcs where the weight changed can be removed.
The augmented graph is now as small as possible.

The CH query algorithm can be reused without modifications.
Another query algorithm is described in~\cite{dsw-cch-15} which does not need priority queues.
It is based on the \emph{elimination tree}.
A node's parent in the elimination tree is its lowest-ranked upward neighbor in the augmented graph.
The ancestors of a node in the elimination tree are exactly the set of nodes that are reachable in a CH search from this node~\cite{bcrw-s-16}.
Thus, instead of exploring the search space through Dijkstra searches, the elimination tree query searches the same nodes by traversing the path to the root in the elimination tree.

\subsection{Shortcut Unpacking Data}\label{sec:shortcut}

The key element of our approach is the information we store with each arc of the augmented graph.
We store time-dependent unpacking information, which allows us to efficiently reconstruct the original path represented by a shortcut for a point in time.
In (C)CH, shortcuts $uv$ are inserted when a node $w$ is contracted and the arcs $uw$ and $wv$ exist.
Thus, a shortcut $uv$ always skips over a triangle $[u,w,v]$.
However, there may be several triangles and which one is the fastest may change over time.
There may also be an arc $uv$ in the input graph which might sometimes be the fastest path.
This is the information our unpacking data structure has to capture.

\begin{figure}
\centering
\begin{tikzpicture}[]
  \node [node] at (0, 0) (u) {$u$};
  \node [node] at (6, 0) (v) {$v$};
  \node [node] at (2, -1.5) (w1) {$w_1$};
  \node [node] at (4, -1.5) (w2) {$w_2$};

  \node at (9, -.6) {
    \begin{tabular}{|c|c|c|}
      \hline
      00:00 & $u w_1$ & $w_1 v$ \\
      \hline
      07:32 & $u w_2$ & $w_2 v$ \\
      \hline
      15:42 & $u w_1$ & $w_1 v$ \\
      \hline
    \end{tabular}
  };

  \draw [->, dashed] (u) -- (v);
  \draw [->] (u) -- (w1);
  \draw [->] (w1) -- (v);
  \draw [->] (u) -- (w2);
  \draw [->] (w2) -- (v);
\end{tikzpicture}
\caption{A shortcut with associated time-dependent expansions.}\label{fig:shortcut_data}
\end{figure}
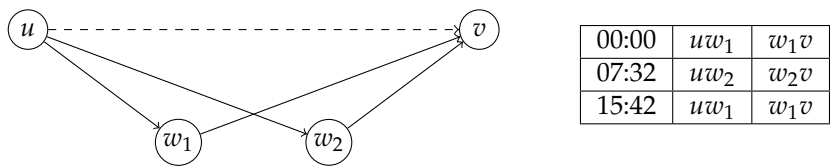

For each arc $uv$, we store a set of time-dependent \emph{expansions} $X_{uv}$ for unpacking.
Figure~\ref{fig:shortcut_data} presents an example.
For an expansion $x \in X_{uv}$, we denote the time during which $x$ represents the shortest path as the \emph{validity interval} $\Pi_x$ of $x$.
When formally referring to the path represented by an expansion, we use the \emph{expand function} $E : X \to V \cup A$.
$E$ either maps to an original arc or to the middle node $w_x$ of the lower triangle $[u, w_x, v]$.
Knowing the middle node for each expansion is also sufficient to obtain longer paths.
These can be computed by unpacking shortcuts recursively.

In our implementation, the expansion information is represented as an array of triples $(\pi, u w_x, w_x v)$.
$\pi$ is the beginning of the validity interval and $uw$ and $wv$ are arc ids.
This information can be stored in 16 bytes for each entry -- 8 bytes for the timestamp and 4 bytes for each arc id.
An expansion can also represent an original arc or no arc at all during a certain time interval.
Both these cases are represented as special arc id values.
Two invalid arc ids indicate no arc at all.
One invalid id indicates that the other arc id represents an original arc.
Beside the expansions, we also maintain a scalar lower bound $\underline{b}_{uv}$ and an upper bound $\overline{b}_{uv}$ on the travel time for each arc.

\begin{algorithm}
\caption{\textsc{LowerTriangleRelax}}\label{alg:lower_triangle_relax}

\DontPrintSemicolon
\SetFuncSty{textsc}

\KwIn{
Preliminary unpacking data for arc $uv$: $X_{uv}, \underline{b}_{uv}, \overline{b}_{uv}$ and travel time function $f_{uv}$.
Travel time functions of lower triangle arcs $f_{uw}$ and $f_{wv}$.}
\KwOut{Improved data $X_{uv}, \underline{b}_{uv}, \overline{b}_{uv}$ and function $f_{uv}$.}
\SetKwFunction{expansion}{NewExpansion}
\BlankLine

$g \gets f_{wv} \circ f_{uw}$\;
$\Pi_{uv} \gets \{ \tau \mid f_{uw}(\tau) \leq g(\tau) \}$\;
$\Pi_{[uwv]} \gets \{ \tau \mid g(\tau) < f_{uw}(\tau) \}$\;
$X_{uv} \gets \{ \expansion{$x, \Pi_{uv} \cap \Pi_x$} \mid x \in X_{uv} \} \cup \{ \expansion{$[u,w,v], \Pi_{[uwv]}$} \}$\;
$f_{uv} \gets \min(f_{uv}, g)$\;
$\underline{b}_{uv} \gets \min(\underline{b}_{uv}, \min_{\tau}(f_{uv}(\tau)))$\;
$\overline{b}_{uv} \gets \min(\overline{b}_{uv}, \max_{\tau}(f_{uv}(\tau)))$\;

\Return{$X_{uv}, \underline{b}_{uv}, \overline{b}_{uv}, f_{uv}$}

\end{algorithm}

During preprocessing, we have to compute the bounds and expansion sets for each arc in the augmented graph.
This is done using the same schema as in CCH.
We iterate over all arcs and relax their lower triangles.
Algorithm~\ref{alg:lower_triangle_relax} depicts the routine for each triangle.
The routine requires travel time functions for all involved arcs.
We maintain these functions during preprocessing but discard them later.
To relax the lower triangle $[u,w,v]$, the functions $f_{uw}$ and $f_{wv}$ are linked and the result is merged with the current function of $f_{uv}$.
Where $[u,v,w]$ is faster, new expansions are inserted into $X_{uv}$.
Where the current $uv$ travel time is faster, the current expansions are kept.
The bounds are updated with the new minimum and maximum of the merged function.

Once the unpacking information for an arc is complete, we can use it to compute the arc's travel time, unpack it to the path in the original graph, or compute the travel time function for the arc.
All these operations follow the same basic schema:
Determine the relevant expansions and apply the operation recursively until arcs from the input graph are reached.
The simplest case is the travel time evaluation.
Algorithm~\ref{alg:eval} depicts this operation.
First, the relevant expansion is determined using binary search.
If it points to an original arc, this arc's travel time can be evaluated and returned.
If the expansion points to a lower triangle, we first recursively evaluate the first arc of the triangle.
Then, the second arc can be evaluated at the entry time plus the travel time of the first arc.

\begin{algorithm}
\caption{\textsc{Eval}}\label{alg:eval}

\DontPrintSemicolon
\SetFuncSty{textsc}

\SetKwFunction{eval}{Eval}

\KwIn{Expansions $X_{uv}$ for edge $uv$, Time $\tau$}
\KwOut{Travel time when traversing $uv$ at $\tau$}
\BlankLine

$x_\tau \gets x \in X_s$ such that $\tau \in \Pi_x$ \tcp{binary search} \;

\eIf{$E(x_\tau) = uv \in A$}{
  \Return{$f_{uv}(\tau)$}\;
}{
  $w_x \gets E(x_\tau)$\;
  $\tau' \gets$ \eval{$X_{uw_x}, \tau$}\;
  \Return{$\tau' +$ \eval{$X_{w_x v}, \tau + \tau'$}}\;
}

\end{algorithm}

Algorithm~\ref{alg:unpack} depicts the procedure for determining the path represented by an expansion set for a given time.
The recursive schema is the same as for \textsc{Eval} but the result is a path instead of a travel time.
Nevertheless, when unpacking lower triangles, we still need to evaluate the first arcs travel time to determine the time for unpacking the second arc.

\begin{algorithm}
\caption{\textsc{Unpack}}\label{alg:unpack}

\DontPrintSemicolon
\SetFuncSty{textsc}

\SetKwFunction{eval}{Eval}
\SetKwFunction{unpack}{Unpack}

\KwIn{Expansions $X_{uv}$ for edge $uv$, Time $\tau$}
\KwOut{Unpacked path $[u,\dots,v]$}
\BlankLine

$x_\tau \gets x \in X_s$ such that $\tau \in \Pi_x$ \tcp{binary search} \;

\eIf{$E(x_\tau) = uv \in A$}{
  \Return{$[u,v]$}\;
}{
  $w_x \gets E(x_\tau)$\;
  $p \gets$ \unpack{$X_{uw_x}, \tau$}\;
  \Return{$p \cdot$ \unpack{$X_{w_x v}, \tau$ $+$ \eval{$p, \tau$}}}\;
}

\end{algorithm}

Constructing the travel time function is also similar and shown in Algorithm~\ref{alg:unpack_profile}.
We recursively unpack expansions until we reach arcs of the original graph where exact travel time functions are available.
We may need to unpack several expansions for different times and combine them.
For each expansion we check if its validity overlaps with the time range for which we want to construct the travel time function.
If so, we recursively retrieve the function for the first arc during this overlap.
Then, we calculate the function for the second arc during the overlap.
For the second arc, the time interval must be shifted by the travel time of the first arc at the start and end of the time interval.
Both functions are then linked and appended to the final function.

Implementing this algorithm naively may cause performance issues since many memory allocations are performed for intermediate results.
We avoid this by keeping all intermediate results in two buffers which are reused for all invocations of this algorithm.
The buffers are stored as dynamically sized arrays (C++ vectors) and can grow dynamically but will never shrink.
Once they have grown to an appropriate size, no more memory allocations will be necessary.
Each buffer can contain many travel time functions stored consecutively.
The link operation will read the last two functions from one buffer and append the result to the other buffer.
Then, the two input functions will be truncated from the first buffer.
After swapping, the buffers can be used again for the next link operation.
Swapping is necessary, because it is not possible to read from and write to the same buffer during the same operation.
The same schema can be employed for joining partial functions (see Figure~\ref{fig:buffers} for a visualization).

\begin{algorithm}[p]
\caption{\textsc{ReconstructTravelTimeFunction}}\label{alg:unpack_profile}

\DontPrintSemicolon
\SetFuncSty{textsc}

\SetKwFunction{profile}{ReconstructTravelTimeFunction}

\KwIn{Expansions $X_{uv}$ for edge $uv$, Time interval $T$}
\KwOut{Exact travel time function $f_{uv}|_{T}$}
\BlankLine

Initialize $f_{uv}$ as function with empty domain\;
\For{$x \in X_{uv}$}{
  $[\tau_x, \pi_x] \gets T \cap \Pi_x$\;
  \eIf{$E(x_\tau) = uv \in A$}{
    $f_x \gets f_{uv}|_{[\tau_x, \pi_x]}$\;
  }{
    $w_x \gets E(x_\tau)$\;
    $f_{u w_x} \gets$ \profile{$u w_x, [\tau_x, \pi_x]$}\;
    $f_{w_x v} \gets$ \profile{$w_x v$, $[\tau_x + f_{u w_x}(\tau_x), \pi_x + f_{u w_x}(\pi_x)]$}\;
    $f_x \gets f_{w_x v} \circ f_{u w_x}$
  }
  $f_{uv} \gets f_{uv} \cup f_x$
}

\Return{$f_{uv}$}

\end{algorithm}

\definecolor{c1}{HTML}{4c72b0}
\definecolor{c2}{HTML}{dd8452}
\definecolor{c3}{HTML}{55a868}
\definecolor{c4}{HTML}{c44e52}
\definecolor{c5}{HTML}{8172b3}
\definecolor{c6}{HTML}{937860}
\definecolor{c7}{HTML}{da8bc3}
\definecolor{c8}{HTML}{ccb974}
\definecolor{c9}{HTML}{64b5cd}

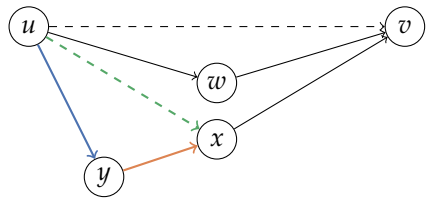
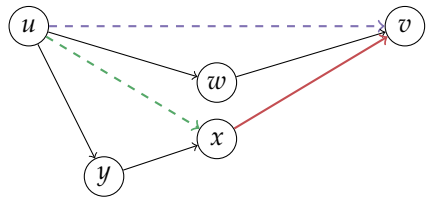
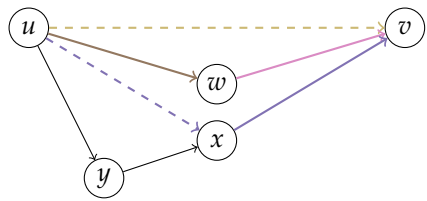
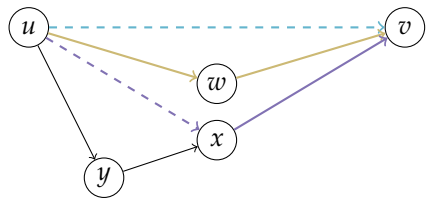
\begin{figure}[p]
\begin{subfigure}{\columnwidth}
\centering
\begin{tikzpicture}[function/.style={rectangle, draw=black, minimum width=1cm, minimum height=.295cm}]
  \node [node] at (0, 0) (u) {$u$};
  \node [node] at (5, 0) (v) {$v$};
  \node [node] at (2.5, -0.75) (w1) {$w$};
  \node [node] at (2.5, -1.5) (w2) {$x$};
  \node [node] at (1, -2) (w3) {$y$};

  \path [->, dashed] (u) edge (v);
  \path [->] (u) edge (w1);
  \path [->] (w1) edge (v);
  \path [->, dashed, draw=c3, thick] (u) edge (w2);
  \path [->] (w2) edge (v);
  \path [->, draw=c1, thick] (u) edge (w3);
  \path [->, draw=c2, thick] (w3) edge (w2);

  \draw (12, -0.2) -- (7, -0.2) -- (7, -.5) -- (12, -.5);
  \draw (12, -1.5) -- (7, -1.5) -- (7, -1.8) -- (12, -1.8);

  \node [function, fill=c1] (f1) at (7.5, -1.65)  {};
  \node [function, fill=c2] (f2) at (8.5, -1.65)  {};
  \node [function, fill=c3] (f3) at (7.5, -0.35)  {};

  \node [circle,inner sep=0.2mm,minimum size=2mm,draw = black] at (8, -1) (op) { + };
  \path [->] (f1) edge (op);
  \path [->] (f2) edge (op);
  \path [->] (op) edge (f3);
\end{tikzpicture}
\caption{Linking $uy$ and $yx$ to compute $ux$.}
\end{subfigure}

\begin{subfigure}{\columnwidth}
\centering
\begin{tikzpicture}[function/.style={rectangle, draw=black, minimum width=1cm, minimum height=.295cm}]
  \node at (0,0.5) {};
  \node [node] at (0, 0) (u) {$u$};
  \node [node] at (5, 0) (v) {$v$};
  \node [node] at (2.5, -0.75) (w1) {$w$};
  \node [node] at (2.5, -1.5) (w2) {$x$};
  \node [node] at (1, -2) (w3) {$y$};

  \path [->, dashed, draw=c5, thick] (u) edge (v);
  \path [->] (u) edge (w1);
  \path [->] (w1) edge (v);
  \path [->, dashed, draw=c3, thick] (u) edge (w2);
  \path [->, draw=c4, thick] (w2) edge (v);
  \path [->] (u) edge (w3);
  \path [->] (w3) edge (w2);

  \draw (12, -0.2) -- (7, -0.2) -- (7, -.5) -- (12, -.5);
  \draw (12, -1.5) -- (7, -1.5) -- (7, -1.8) -- (12, -1.8);

  \node [function, fill=c3] (f3) at (7.5, -0.35)  {};
  \node [function, fill=c4] (f4) at (8.5, -0.35)  {};
  \node [function, fill=c5] (f5) at (7.5, -1.65)  {};

  \node [circle,inner sep=0.2mm,minimum size=2mm,draw = black] at (8, -1) (op) { + };
  \path [->] (f3) edge (op);
  \path [->] (f4) edge (op);
  \path [->] (op) edge (f5);
\end{tikzpicture}
\caption{Linking $ux$ and $xv$ to compute a part of $uv$ with inverted buffer roles.}
\end{subfigure}

\begin{subfigure}{\columnwidth}
\centering
\begin{tikzpicture}[function/.style={rectangle, draw=black, minimum width=1cm, minimum height=.295cm}]
  \node at (0,0.5) {};
  \node [node] at (0, 0) (u) {$u$};
  \node [node] at (5, 0) (v) {$v$};
  \node [node] at (2.5, -0.75) (w1) {$w$};
  \node [node] at (2.5, -1.5) (w2) {$x$};
  \node [node] at (1, -2) (w3) {$y$};

  \path [->, dashed, draw=c8, thick] (u) edge (v);
  \path [->, draw=c6, thick] (u) edge (w1);
  \path [->, draw=c7, thick] (w1) edge (v);
  \path [->, dashed, draw=c5, thick] (u) edge (w2);
  \path [->, draw=c5, thick] (w2) edge (v);
  \path [->] (u) edge (w3);
  \path [->] (w3) edge (w2);

  \draw (12, -0.2) -- (7, -0.2) -- (7, -.5) -- (12, -.5);
  \draw (12, -1.5) -- (7, -1.5) -- (7, -1.8) -- (12, -1.8);

  \node [function, fill=c6] (f6) at (7.5, -0.35)  {};
  \node [function, fill=c7] (f7) at (8.5, -0.35)  {};
  \node [function, fill=c5] (f5) at (7.5, -1.65)  {};
  \node [function, fill=c8] (f8) at (8.5, -1.65)  {};

  \node [circle,inner sep=0.2mm,minimum size=2mm,draw = black] at (8, -1) (op) { + };
  \path [->] (f6) edge (op);
  \path [->] (f7) edge (op);
  \path [->] (op) edge (f8);
\end{tikzpicture}
\caption{Linking $uw$ and $wv$ to compute the other part of $uv$.}
\end{subfigure}

\begin{subfigure}{\columnwidth}
\centering
\begin{tikzpicture}[function/.style={rectangle, draw=black, minimum width=1cm, minimum height=.295cm}]
  \node at (0,0.5) {};
  \node [node] at (0, 0) (u) {$u$};
  \node [node] at (5, 0) (v) {$v$};
  \node [node] at (2.5, -0.75) (w1) {$w$};
  \node [node] at (2.5, -1.5) (w2) {$x$};
  \node [node] at (1, -2) (w3) {$y$};

  \path [->, dashed, draw=c9, thick] (u) edge (v);
  \path [->, draw=c8, thick] (u) edge (w1);
  \path [->, draw=c8, thick] (w1) edge (v);
  \path [->, dashed, draw=c5, thick] (u) edge (w2);
  \path [->, draw=c5, thick] (w2) edge (v);
  \path [->] (u) edge (w3);
  \path [->] (w3) edge (w2);

  \draw (12, -0.2) -- (7, -0.2) -- (7, -.5) -- (12, -.5);
  \draw (12, -1.5) -- (7, -1.5) -- (7, -1.8) -- (12, -1.8);

  \node [function, fill=c9] (f9) at (7.5, -0.35)  {};
  \node [function, fill=c5] (f5) at (7.5, -1.65)  {};
  \node [function, fill=c8] (f8) at (8.5, -1.65)  {};

  \node [circle,inner sep=0.2mm,minimum size=2mm,draw = black] at (8, -1) (op) { + };
  \path [->] (f5) edge (op);
  \path [->] (f8) edge (op);
  \path [->] (op) edge (f9);
\end{tikzpicture}
\caption{Combining both parts of $uv$ into one function which yields the final result.}
\end{subfigure}
\caption{Avoiding allocations when reconstructing shortcut travel time functions with two reusable buffers.}
\label{fig:buffers}
\end{figure}

\subsection{Preprocessing}\label{sec:preprocessing}

In this section, we present our preprocessing algorithms.
The first phase of CCH preprocessing is performed only on the topology of the graph.
Since no travel time functions are involved, we can adapt the algorithms of \citet{dsw-cch-15} without modification.
We use InertialFlowCutter~\cite{ghuw-fbndocch-19} to obtain the nested dissection order.
To generate the augmented graph, we implement an improved contraction algorithm first presented in~\cite{z-cchtc-19}.
When contracting a node, we insert all upward neighbors of the current node only into the neighborhood of its lowest ranked upward neighbor.
This algorithm can be implemented to run in linear time in the size of the output graph.

The goal of the second phase of preprocessing -- the \emph{customization} -- for classical CCH is to compute the shortcut weights.
For our approach, we have to compute the travel time bounds and time-dependent expansions for all arcs in the augmented graph.
Recall that a shortcut $u v$ always bypasses one or many lower triangles $[u,w_i,v]$ for different nodes $w_i$, where $w_i$ has lower rank than $u$ and $v$.
For the bounds, we want to find the minimum and maximum travel time of the fastest travel time function between $u$ and $v$ over any $w_i$.
For the expansions, we need to determine for each point in time which lower triangle is the fastest.
Assuming we know the final travel time functions of all $u w_i$ and $w_i v$, we can compute this using the \textsc{LowerTriangleRelax} algorithm (see Algorithm~\ref{alg:lower_triangle_relax}).
This leads to the following algorithmic schema:
Maintain a set of necessary travel time functions in memory starting with the functions from the input graph.
Iterate over all arcs in the augmented graph in a bottom-up fashion.
For each arc enumerate lower triangles.
Link and merge their functions to compute the function, bounds, and expansions of the current arc.
Keep the current arc's travel time function in memory until it is no longer needed.

We implement this schema as follows:
We process all arcs $u v$ ordered ascending by their lower ranked endpoint.
Since the middle node $w$ of a lower triangle $[u,w,v]$ has always lower rank than $u$ and $v$, the arcs $u w$ and $w v$ will have been processed already.
To process an arc $u v$ we enumerate lower triangles $[u,w,v]$.
Perform \textsc{LowerTriangleRelax} for each triangle in both directions.
Once all arcs $u v$ have been processed where $u$ is the lower ranked endpoint, we drop the travel time functions of all arcs $w u$ where $u$ is the higher ranked endpoint.
Algorithm~\ref{alg:customization} depicts this in pseudo code.

\begin{algorithm}
\caption{\textsc{Customization}}\label{alg:customization}

\DontPrintSemicolon
\SetFuncSty{textsc}

\SetKwFunction{ltrelax}{RelaxLowerTriangle}
\SetKwFunction{drop}{Drop}

\KwIn{Augmented graph $G = (V, A)$ with travel time functions for input graph arcs, node ranking $r$}
\KwOut{Expansion data and bounds for all arcs}
\BlankLine

\For{$u \in V$ ordered by $r(u)$}{
\For{$uv \in \{ uv \mid uv \in A, r(u) < r(v) \}$}{
\For{$w \in \{ w \mid uw \in A, wv \in A, r(w) < r(u) \}$}{
\ltrelax{$[u,w,v], f_{uv}, f_{uw}, f_{wv}$}\;
\ltrelax{$[v,w,u], f_{vu}, f_{vw}, f_{wu}$}\;
}
}
\For{$uw \in \{ uw \mid uw \in A, r(w) < r(u) \}$}{
\drop{$f_{uw}$}\;
\drop{$f_{wu}$}\;
}
}

\end{algorithm}

When enumerating triangles, we order them ascending by $\underline{b}_{u w} + \underline{b}_{w v}$.
This way, we process triangles first, which are likely faster.
This gives us preliminary bounds on the travel time of $u v$.
Before linking the functions of another triangle $f_{u w}$ and $f_{w v}$, we check if $\overline{b}_{u v} \leq \underline{b}_{u w} + \underline{b}_{w v}$.
If so, the linked path would be dominated by the shortcut, and we can skip linking and merging completely.
If not, we link $f_{u w}$ and $f_{w v}$ and obtain $f_{[u,w,v]}$.
We still can skip merging if one function is strictly smaller than the other, that is either $\overline{b}_{u v} \leq \min(f_{[u,w,v]})$ or $\max(f_{[u,w,v]}) \leq \underline{b}_{u v}$.
Even if the bounds overlap, one function might still dominate the other.
To check for this case, we simultaneously sweep over the breakpoints of both functions, determining the value of the respectively other function by linear interpolation.
Only when this check fails, we perform the merge operation.

Before the time-dependent customization, we first use the basic and perfect customization algorithms from~\cite{dsw-cch-15} to compute preliminary scalar upper and lower bounds for all arcs.
With these bounds, we can skip additional linking and merging operations.
Employing perfect customization, we can remove some arcs completely, when a dominating path through higher ranked nodes exists.

\subparagraph*{Parallelization}

We employ both loop based and task based parallelism. 
The original CCH publication~\cite{dsw-cch-15} suggests processing arcs with their lower ranked endpoint on the same \emph{level} in parallel.
The level of a node is the maximum level of its downward neighbors increased by one, or zero if the node does not have downward neighbors.
We use this approach to process arcs in the top-level separators.
However, this approach requires a synchronization step on each level.
This is detrimental to load balancing.
Thus, we use a different strategy when possible.

In~\cite{bsw-rttau-19}, a task based parallelism approach utilizing the separator decomposition of the graph is proposed.
Each task is responsible for a subgraph $G'$.
Removing the top-level separator in $G'$ decomposes the subgraph into two or more disconnected components.
For each component, a new task is spawned to process the arcs the component.
After all child tasks are completed, the arcs in the separator are processed utilizing the loop based parallelization schema.
If the size of subgraph $G'$ is below a certain threshold, the task processes the arcs in $G'$ sequentially without spawning subtasks.
We use $n/(\alpha \cdot c)$ as the threshold with $c$ being the number of cores and the tuning parameter $\alpha$ set to $32$ as suggested by \citet{bsw-rttau-19}.

\subparagraph*{Approximation}

As we process increasingly higher ranked arcs, the associated travel time functions become increasingly complex.
This leads to two problems.
First, linking and merging becomes very time-consuming as running times scale with the complexity of the functions.
Second, storing these functions -- even though it is only temporary -- requires a lot of memory.
We employ approximation to mitigate these issues.
However, for exact queries, we need exact unpacking information.
We achieve this by lazily reconstructing parts of exact travel time functions during merging.

When approximating, we do not store one approximated function but two -- a lower bound function and an upper bound function with maximum difference $\epsilon$ where $\epsilon$ is a configurable parameter.
These approximations replace the exact function stored for later merge operations and will also be dropped when no longer needed.
To obtain the bound functions, we first compute an approximation using the algorithm of Douglas and Peucker~\cite{dp-arnpr-73}\footnote{
Previous works~\cite{bgsv-mtdtt-13,bdpw-dtdrp-16} have reported using the algorithm of Imai and Iri~\cite{ii-a-87} for approximation.
Given a maximum error bound $\epsilon$, this algorithm can compute in linear time the piecewise linear function with the minimum number of breakpoints within the given bound.
The Douglas--Peucker algorithm has a quadratic worst case running time and no such guarantees on the number of breakpoints in the approximation.
However, the theoretic guarantees of the Imai--Iri algorithm come at the cost of considerable implementation complexity and high constant runtime factors.
Preliminary experiments showed that, compared to Imai--Iri, our Douglas--Peucker implementation actually produces insignificantly more breakpoints and also runs faster due to better constants.
In addition, the implementation needs 30 instead of 400 lines of code, so we use the Douglas--Peucker variant.
}.
Then, we add or subtract $\epsilon$ to the value of each breakpoint to obtain an upper or lower bound, respectively.
This bounds are valid, but they may not be as tight as possible.
Therefore, we iterate over all approximated points and move each point back towards the original function.
Both adjacent segments in the approximated functions have a minimum absolute error to the original function.
We move the breakpoint by the smaller of the two errors.
This yields sufficiently good bounds.

When linking approximated functions, we link both lower and both upper bound functions.
Linking two lower bounds yields a valid lower bound of the linked exact functions because of the FIFO property.
The same argument holds for upper bounds.

\begin{figure}
\centering
\includegraphics[width=.5\columnwidth]{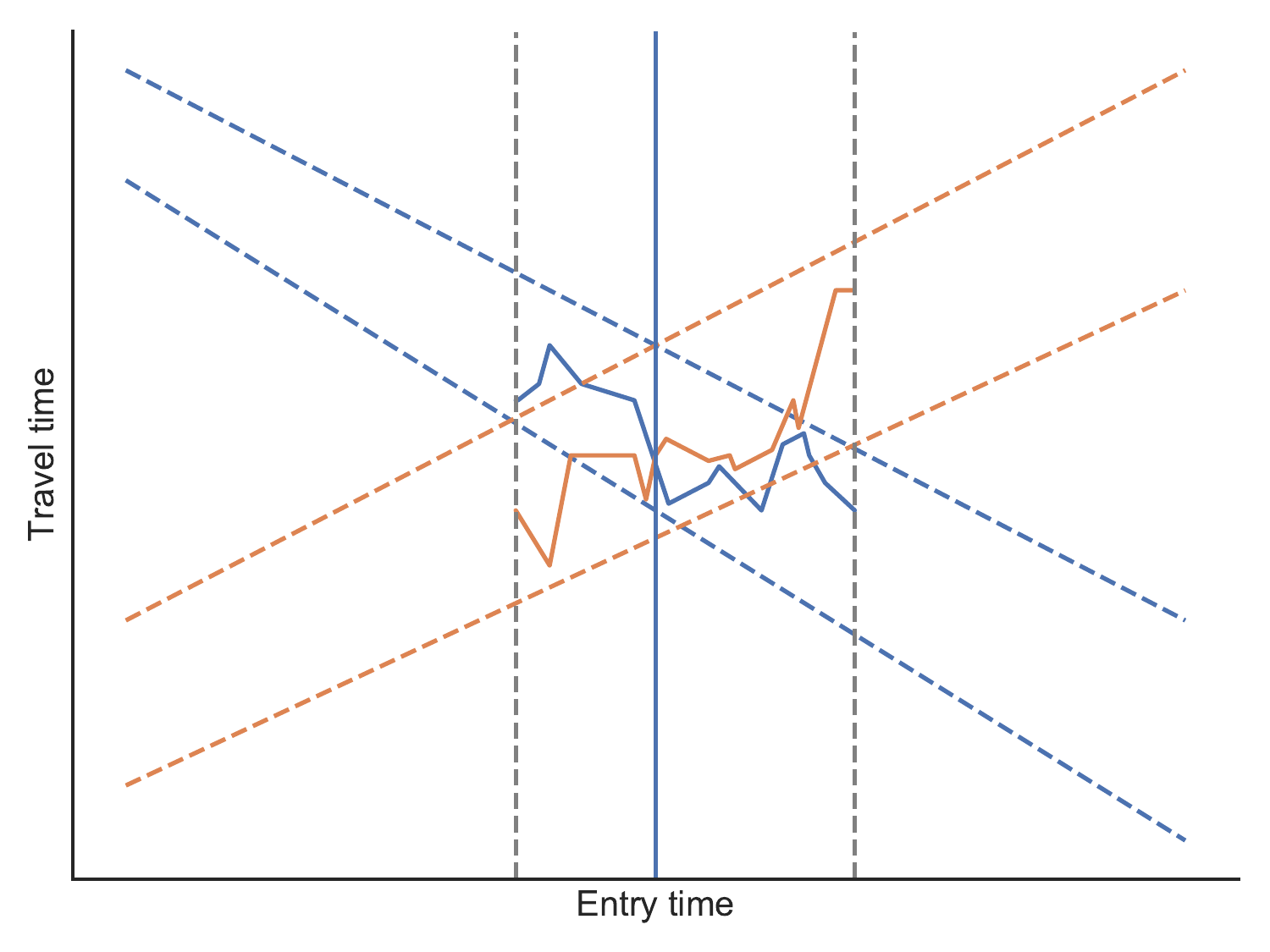}
\caption{Merging approximated travel time functions by reconstructing the exact functions where bounds overlap.}\label{fig:merge_approx}
\end{figure}

Merging approximated functions is slightly more involved.
Our goal is to determine the exact expansions for each arc.
We use the approximated bounds to narrow down the time ranges where intersections are possible.
To identify these parts, we merge the first function's lower bound with the second function's upper bound and vice versa.
Where the bounds overlap, an intersection might occur.
We then obtain the exact functions in the overlapping parts using Algorithm~\ref{alg:unpack_profile} and perform the exact merging.
To obtain approximated upper and lower bounds of the merged function, we merge both lower bounds and both upper bounds (see Figure~\ref{fig:merge_approx} for a visualization).

We approximate whenever a function has more than $\beta$ breakpoints after merging.
This includes already approximated functions.
Both $\beta$ and the maximum difference $\epsilon$ are tuning parameters which influence the performance (but not the correctness).
We evaluate different choices in Section~\ref{sec:exp:customization}.

\subsection{Queries}\label{sec:queries}
\unskip
\subsubsection{Earliest Arrival Queries}

Recall that, for an earliest arrival query, we are given a source node $s$, a target $t$ and a departure time $\tau$ from $s$.
The goal is to obtain the earliest arrival at $t$.
Compared to a standard (C)CH query, our query algorithm has to deal with two challenges.
First, we cannot simply perform a backwards search, as we do not know the arrival time at the target node.
Second, to evaluate the travel time of a shortcut, we need to obtain the path in the original graph which is an expensive operation.
To address the first challenge, the query is split in two phases.
In the first phase we obtain a subgraph on which we can run a forward-only Dijkstra-like search in the second phase.
We now present the basic query algorithm and later introduce optimizations to address the second challenge.

In the first phase, the union of the subgraphs reachable through arcs to higher ranked neighbors from $s$ and $t$ is obtained.
We construct these subgraphs by traversing the elimination tree starting from both $s$ and $t$ to the root and marking all encountered arcs as part of the search space.
The backward search from $t$ maintains parent pointers to represent the subgraph:
For each encountered arc $uv$ (where $v$ has the higher rank), we store the arc id and the tail $u$ at $v$.
This allows efficiently traversing these downward arcs in the forward-only Dijkstra in the second phase.
In the second phase, we run Dijkstra's algorithm on the combined search spaces.
Shortcut travel times are evaluated with Algorithm~\ref{alg:eval}.

By the construction of CH, the search space contains the shortest path, thus Dijkstra's algorithm will find it and this algorithm will determine the correct earliest arrival at $t$.
However, the search space is bigger than strictly necessary.
This slows down the query.
In the next paragraph, we discuss how to construct smaller subgraphs using an elimination tree interval query.

\subparagraph*{Elimination Tree Interval Query}

The elimination tree interval query is a bidirectional search starting from both the source node $s$ and the target node $t$.
It constructs a smaller subgraph for the second phase.
We denote this subgraph as a \emph{shortest path corridor}.
Node labels contain an upper $\overline{t}_v$ and a lower bound $\underline{t}_v$ on the travel time to/from the search origin and a parent pointer to the previous node and the respective arc id.
The bounds $\overline{t}_s$, $\overline{t}_t$, $\underline{t}_s$, $\underline{t}_t$ are all initialized to zero in their respective direction, all other bounds to infinity.
We also track tentative travel time bounds for the total travel time from $s$ to $t$.
For both directions, the path from the start node to the root of the elimination tree is traversed.
For each node $u$, all arcs $u v$ to higher ranked neighbors are relaxed, that is checking if $\overline{t}_u + \overline{b}_{u v} < \overline{t}_v$ or $\underline{t}_u + \underline{b}_{u v} < \underline{t}_v$ and improving the bounds of $v$ if possible.
When the new travel time bounds from an arc relaxation overlap with the current bounds, more than one label has to be stored.
As an optimization~\cite{bsw-rttau-19}, nodes can be skipped if the lower bound on the travel time to it is already greater than the tentative upper bound on the total travel time between $s$ and $t$.
After both directions are finished, we have several nodes in the intersection of the search spaces.
Where the sum of the forward and backward distance bounds of such a node overlaps with the total travel time bounds, the parent pointers are traversed to the search origin and all arcs on the paths are marked as part of the shortest path corridor.

\subparagraph*{Lazy Shortcut Unpacking}

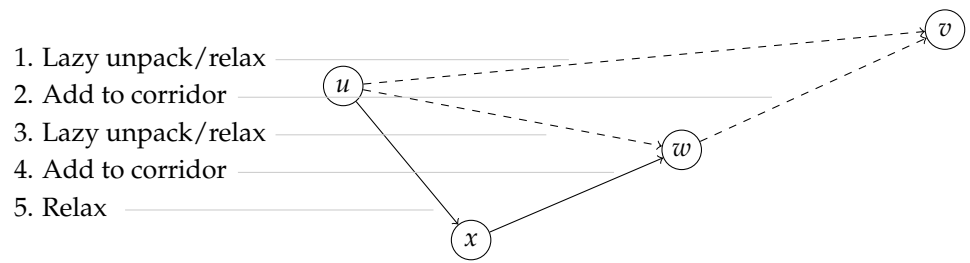
\begin{figure}
\centering
\begin{tikzpicture}[]
  \node [node] at (0, -.75)   (u) {$u$};
  \node [node] at (8, 0)      (v) {$v$};
  \node [node] at (4.5, -1.6) (w) {$w$};
  \node [node] at (1.7, -2.8) (x) {$x$};

  \draw [->, dashed] (u) to (v);
  \draw [->, dashed] (u) to (w);
  \draw [->, dashed] (w) to (v);
  \draw [->] (u) to (x);
  \draw [->] (x) to (w);

  \node [minimum width=1cm, minimum height=.5cm, anchor=north west] at (-4.5, .4 -.5)  (text_uv) {1. Lazy unpack/relax};
  \node [minimum width=1cm, minimum height=.5cm, anchor=north west] at (-4.5, .4 -1)   (text_wv) {2. Add to corridor};
  \node [minimum width=1cm, minimum height=.5cm, anchor=north west] at (-4.5, .4 -1.5) (text_uw) {3. Lazy unpack/relax};
  \node [minimum width=1cm, minimum height=.5cm, anchor=north west] at (-4.5, .4 -2)   (text_xw) {4. Add to corridor};
  \node [minimum width=1cm, minimum height=.5cm, anchor=north west] at (-4.5, .4 -2.5) (text_vx) {5. Relax};

  \draw [draw=black!20] ( -.9, .1 -.5)  -- (3,   .1 -.5);
  \draw [draw=black!20] (-1.4, .1 -1)   -- (5.7,   .1 -1);
  \draw [draw=black!20] ( -.9, .1 -1.5) -- (2.7, .1 -1.5);
  \draw [draw=black!20] (-1.4, .1 -2)   -- (3.6, .1 -2);
  \draw [draw=black!20] (-2.9, .1 -2.5) -- (1.2, .1 -2.5);
\end{tikzpicture}
\caption{
Lazy relaxation of arc $u v$.
Since $u v$ is a shortcut, it needs to be unpacked.
This causes $w v$ to be added to the corridor and $u w$ to be relaxed.
Relaxing $u w$ causes $x w$ to be added to the corridor and $u x$ to be relaxed.
In this example, $u x$ is an original arc and the recursion stops.
$x w$ will be relaxed (or unpacked) only once $x$ is popped from the queue.
}\label{fig:lazy_unpack}
\end{figure}

In the second query phase, we perform Dijkstra's algorithm on the corridor obtained in the first phase.
In the basic query, shortcuts are unpacked completely to evaluate their travel time.
However, this may cause unnecessary and duplicate unpacking work.
We now describe an optimized version of the algorithm which performs unpacking lazily.
The algorithm starts with the same initialization as a regular TD-Dijkstra.
All earliest arrivals are set to infinity, except for the start node which is set to the departure time.
The start node is inserted into the queue.
Then, nodes are popped from the queue until it is empty or the target node is reached.
For each node, all outgoing arcs within the shortest path corridor are relaxed.
When an arc is from the input graph, its travel time function can be evaluated directly.
Shortcut arcs, however, need to be unpacked.
The lazy unpacking algorithm defers as much work as possible:
Only the first arc of the triangle of each shortcut will be recursively unpacked until an input arc is reached, the second arc will be added to the corridor.
Figure~\ref{fig:lazy_unpack} shows an example.
This way, we unpack only the necessary parts and avoid relaxing arcs multiple times when shortcuts share the same paths.

\subparagraph*{Corridor A*}

The query can be accelerated further, by using the lower bounds obtained during the elimination tree interval query as potentials for an A*-search.
For nodes in the CH search space of $t$, the lower bounds from the backward search can be used.
For nodes in the CH search space of $s$, we start at the meeting nodes from the corridor search and propagate the bounds backwards down along the parent pointers.
This yields potentials for all nodes in the initial corridor.
However, we also need potentials for nodes added to the corridor through unpacking.
These potentials are computed during the shortcut unpacking.
When unpacking a shortcut $u v$ into the arcs $u w$ and $w v$, then $\rho(w)$ will be set to $\min(\rho(w), \rho(v) + \underline{b}_{w v})$.

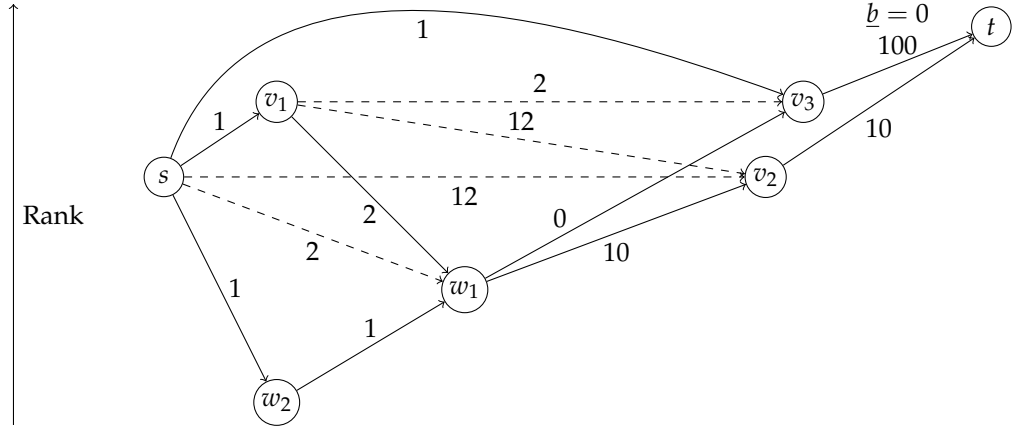
\begin{figure}
\centering
\begin{tikzpicture}[gray/.style={draw = black!25}]
  \path [->] (-2, -3.3) edge node [midway, right] {Rank} (-2, 2.3);

  \node [node] at (0, 0) (s) {$s$};
  \node [node] at (1.5, 1) (v1) {$v_1$};
  \node [node] at (8, 0) (v2) {$v_2$};
  \node [node] at (8.5, 1) (v3) {$v_3$};
  \node [node] at (11, 2) (t) {$t$};

  \node [node] at (4, -1.5) (w1) {$w_1$};
  \node [node] at (1.5, -3) (w2) {$w_2$};

  \path [->] (s) edge node [midway, right] {1} (w2);
  \path [->, dashed] (s) edge node [midway, below] {2} (w1);
  \path [->] (s) edge node [midway, above] {1} (v1);
  \path [->, dashed] (s) edge node [midway, below] {12} (v2);
  \path [->] (s) edge[out=70, in=160] node [midway, below] {1} (v3);

  \path [->, dashed] (v1) edge node [midway, above] {12} (v2);
  \path [->, dashed] (v1) edge node [midway, above] {2} (v3);
  \path [->] (v1) edge node [midway, below] {2} (w1);

  \path [->] (v2) edge node [midway, below=.1cm] {10} (t);
  \path [->] (v3) edge node [midway, above, align=center] {$\underline{b} = 0$ \\ 100} (t);

  \path [->] (w2) edge node [midway, above] {1} (w1);
  \path [->] (w1) edge node [midway, below] {10} (v2);
  \path [->] (w1) edge node [near start, above] {0} (v3);

\end{tikzpicture}
\caption{
Example of a query where our A* potentials lead to a non-label-setting query.
Dashed arcs are shortcuts.
The shortcut weights are not known to the query algorithm.
}\label{fig:astar_non_label_setting}
\end{figure}

Justifying that A* with these potentials will always find the correct earliest arrival is surprisingly non-trivial.
In fact, these potentials are not \emph{feasible} in the sense that $\forall \tau \in T, uv \in A : f_{uv}(\tau) - \rho(u) + \rho(v) \geq 0$~\cite{gh-cspas-05}.
Figure~\ref{fig:astar_non_label_setting} shows an example where the term becomes negative and the same node has to be popped several times from the queue.
Assume that all arcs have a constant travel time for the departure time of this query and lower bounds are equal to the travel time.
The exception is $v_3 t$ which has constant travel time 100 during this query but the lower bound is zero.
We use zero weights to simplify the example.
They are not strictly necessary for such an example.
The shortest path from $s$ to $t$ is $[s, w_2, w_1, v_2, t]$ and has length 22.
After $s$ is settled, the queue will contain $v_3$ with key $1+0$ (distance plus lower bound to $t$), $v_1$ with key $1+2$ and $w_2$ with key $1 + 21$.
Then, $v_3$ will be settled which will insert $t$ with key $101+0$ into the queue.
Then, $v_1$ will be settled and $w_1$ will be inserted into the queue with key $3+0$.
Then, $w_1$ will be settled even though the current distance of 3 is greater than the actual shortest distance of 2.
This will insert $v_2$ with key $13 + 10$ into the queue.
Now, $w_2$ will be popped and the distance to $w_1$ will be improved and it will be reinserted into the queue with key $2+20$.
$w_1$ will be popped immediately afterwards which improves the distance and key of $v_2$ which is the next node to be popped from the queue.
After it has been processed, the final distance to $t$ (22) is known, and $t$ is the final node to be settled.

We claim that once $t$ is popped from the queue the algorithm \emph{always} has found the correct earliest arrival.
The reason is that for all nodes $v$ on the shortest path $\operatorname{ea}_v + \rho(v) \leq \operatorname{ea}_t$ holds.
Since this is the queue key, all nodes on the shortest path will have been popped (possibly several times) before $t$.
Let $P = [s, \dots, t]$ be the desired shortest path from $s$ to $t$ when departing from $s$ at $\tau$ and $P[v..] \subset P$ the subpath from a node $v \in P$ to $t$.
We denote the global lower bound on the travel time between a node $v$ and $t$ by $\underline{f}_{vt}$ and the lower bound travel time on $P[v..]$ by $\underline{f}_P[v..]$.
For nodes in the initial corridor obtained by the interval query, $\operatorname{ea}_v + \rho(v) \leq \operatorname{ea}_t$ always holds because the potential $\rho(v)$ is set to the global lower bound $\underline{f}_{vt}$.
However, nodes $u$ added later to the corridor through unpacking may have a greater potential than $\underline{f}_{ut}$ as depicted in the example.
However, their final potential cannot be greater than the lower bound of the travel time on the shortest path $\underline{f}_P[u..]$.
This is enough to satisfy $\operatorname{ea}_u + \rho(u) \leq \operatorname{ea}_t$.
In addition, the potential value will be set to this final value before $t$ is settled.
Assume for contraction that $p_i$ is the first node on $P$ for which this statement does not hold.
$p_i$ cannot be a node from the initial corridor.
However, $\rho(p_i)$ will be set at most to $\underline{f}_P[p_{i}..]$ once $p_{i-1}$ is settled which by assumption happens before $t$ is settled.
This is a contraction.
Thus, $\operatorname{ea}_v + \rho(v) \leq \operatorname{ea}_t$ holds for all nodes $v \in P$ and the query algorithm always finds the correct earliest arrival when terminating once $t$ is popped from the queue.

\subsubsection{Profile Queries}

A profile query asks for the function of the fastest travel time between two nodes over a given time period $T$.
Without loss of generality, we assume that the $T$ equals the entire time period covered by the input network.
As discussed in Section~\ref{sec:prelim}, such a query can be answered with TD-Profile-Dijkstra.
However, TD-Profile-Dijkstra exhibits both prohibitive running time and memory consumption.
Consider a path $[v_0,\dots,v_k]$ where the travel time function of each arc has $b$ breakpoints.
In general, linking two functions $f$ and $g$ creates a new function with $\Theta(|f|+|g|)$ breakpoints.
Thus, the travel time function from $v_0$ to node $v_i$ contains $\Theta(i \cdot b)$ breakpoints.
When applying TD-Profile-Dijkstra to compute the function between $v_0$ and $v_k$, the total memory consumption and the running time grows quadratically with the length of the path.
We conclude that Dijkstra-based approaches to profile queries are not a promising direction.
The experiments with Dijkstra-based TCH profile queries in~\cite{bgsv-mtdtt-13} support this conclusion.
We also performed preliminary experiments with a proof-of-concept implementation where we adapted our earliest arrival query algorithm to the profile query setting.
The Dijkstra-based approach was more than an order of magnitude slower than the approach described in the following. 
Instead of a Dijkstra-like search, our algorithm uses contraction and methods from the preprocessing to construct the desired profile.

Our algorithm has four phases.
The first phase uses the elimination tree interval query to obtain a shortest path corridor and is the same as for earliest arrival queries.
In the second phase, we obtain travel time functions for all arcs in the shortest path corridor.
During the third phase, additional shortcuts will be inserted and their unpacking data will be computed, reusing the preprocessing algorithms.
The result is a new $st$ shortcut with exact expansions $X_{st}$.
From this unpacking information, the exact travel time function (among other things) can be obtained in the fourth phase.
We now describe Phases 2--4 in detail.

\subparagraph*{Reconstruction}
In this phase, we obtain travel time functions for all arcs in the shortest path corridor.
Similar to the customization, the obtained travel time functions may be either exact, or approximated upper and lower bound functions.
This keeps the memory consumption low and linking and merging operations fast.
We obtain these functions by first recursively reconstructing the travel time functions of all arcs referenced by expansions.
The reconstructed functions will be saved for both arcs in the corridor and unpacked arcs, in case another reconstruction operation might reuse an arc.
If all arcs referenced by the expansions have an exact function available, we can compute an exact travel time function for the current arc.
If not, we end up with an approximation.
After reconstruction, we check if a function has more than $\beta$ breakpoints.
If that is the case, we approximate it to reduce the complexity (see Section~\ref{sec:preprocessing}).

\subparagraph*{Contraction}

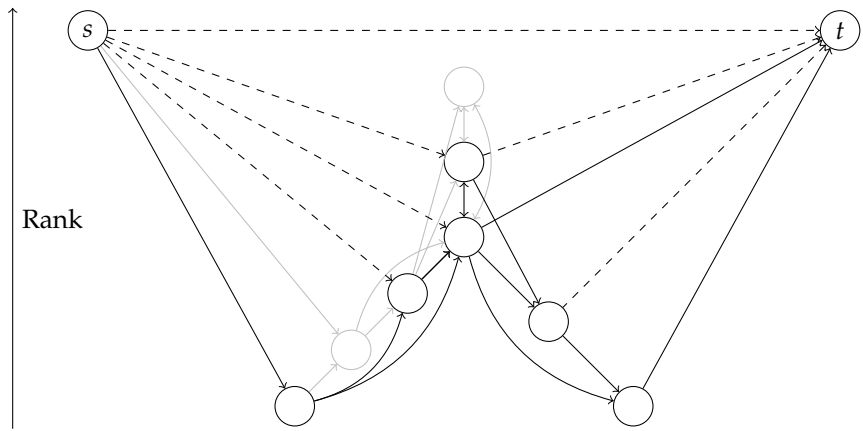
\begin{figure}
\centering
\begin{tikzpicture}[gray/.style={draw = black!25}]
  \node [node] at (0, 0) (s) {$s$};
  \node [node] at (10, 0) (t) {$t$};

  \node [node, gray] at (5, -0.75) (r) {};
  \node [node] at (5, -1.75) (v1) {};
  \node [node] at (5, -2.75) (v2) {};

  \node [node] at (4.25, -3.5) (s1) {};
  \node [node, gray] at (3.5, -4.25) (s2) {};
  \node [node] at (2.75, -5) (s3) {};

  \node [node] at (6.125, -3.875) (t1) {};
  \node [node] at (7.25, -5) (t2) {};

  \path [->] (-1, -5.3) edge node [midway, right] {Rank} (-1, 0.3);

  \draw [->, dashed] (s) -- (t);
  \draw [->, gray] (s3) -- (s2);
  \draw [->, gray] (s2) -- (s1);
  \draw [->] (s1) -- (v2);
  \draw [<-] (t2) -- (t1);
  \draw [<-] (t1) -- (v2);
  \draw [->] (s1) -- (v2);
  \draw [<->] (v2) -- (v1);
  \draw [<->, gray] (v1) -- (r);
  \path [<->, gray] (v2) edge[bend right=30] (r);
  \path [->] (v1) edge (t1);
  \path [->, gray] (s1) edge (v1);
  \path [->, gray] (s1) edge (r);
  \path [->, gray] (s2) edge[bend left] (v2);

  \path [->] (s3) edge[bend right] (s1);
  \path [->] (s3) edge[bend right] (v2);
  \path [->] (v2) edge[bend right] (t2);

  \draw [->] (s) -- (s3);
  \draw [->, gray] (s) -- (s2);
  \draw [->, dashed] (s) -- (s1);
  \draw [->, dashed] (s) -- (v2);
  \draw [->, dashed] (s) -- (v1);

  \draw [<-] (t) -- (t2);
  \draw [<-, dashed] (t) -- (t1);
  \draw [<-] (t) -- (v2);
  \draw [<-, dashed] (t) -- (v1);

\end{tikzpicture}
\caption{
Example of profile query search space with inserted shortcuts.
Gray nodes and arcs are not in the shortest path corridor.
Dashed arcs are new shortcuts.
}\label{fig:profile_query_shortcuts}
\end{figure}

In the third phase, we insert additional shortcuts and compute their unpacking data by simulating the contraction of the nodes in the shortest path corridor.
We reuse the existing nested dissection order.
The ranks of $s$ and $t$ are increased such that they are higher in the hierarchy than all other nodes.
This construction leads to a shortcut between $s$ and $t$, one between $s$ and each node in the corridor on the path from $s$ to the elimination tree root, and one from each node in the corridor on the path from $t$ to the root to $t$.
Some of these shortcuts may already exist
See Figure~\ref{fig:profile_query_shortcuts} for an illustration.

These new shortcuts are now processed as in Algorithm~\ref{alg:customization} to compute their unpacking data.
We initialize the shortcut bounds with the bounds obtained from the elimination tree query.
This allows to prune unnecessary operations.
We process shortcuts ordered by their lower ranked endpoint.
For each shortcut we enumerate and relax lower triangles using Algorithm~\ref{alg:lower_triangle_relax}.
We can enumerate these triangles efficiently using the parent pointer from the interval query.
Each shortcut has an endpoint node in the corridor.
The parent pointers of this node correspond to the triangles that need to be relaxed.
The shortcuts from $s$ and the shortcuts to $t$ are independent of each other and can be processed in parallel.
The lower triangles of the $st$ shortcut can be enumerated by iterating over the meeting nodes from the interval query.
We also employ the triangle sorting optimization.

\subparagraph*{Extraction}

In this final phase, we can use the unpacking information of the $st$ shortcut to efficiently compute the final result.
The shortcut already contains a possibly approximated travel time function from the contraction phase.
This may suffice for some applications.
If the shortcut contains only an approximation, but we need an exact travel time profile, we can use Algorithm~\ref{alg:unpack_profile} to compute it.
For some practical applications, the different shortest paths over the day may be more useful than the travel time profile.
Algorithm~\ref{alg:unpack_paths} depicts a routine to compute path switches and the associated shortest paths.
The algorithm follows the same schema as all unpacking algorithms.
The operation is recursively applied to all expansions limited to the validity time of the expansion.
Only the \textsc{Combine} operation is more involved.
It performs a coordinated linear sweep over the path sets from $u w_x$ and $w_x v$ and appends the paths where the validity intervals overlap.
For the paths from $w_x$ to $v$, we only know the validity times with respect to departure at $w_x$.
To obtain the corresponding departure time at $u$, we reverse evaluate the current $u w_x$ path, i.e., we successively evaluate the inverted arrival time function of all arcs on the path in reverse order.

\begin{algorithm}
\caption{\textsc{UnpackPaths}}\label{alg:unpack_paths}

\DontPrintSemicolon
\SetFuncSty{textsc}

\SetKwFunction{unpack}{UnpackPaths}
\SetKwFunction{combine}{Combine}

\KwIn{Expansions $X_{uv}$ for edge $uv$, Time interval $T$}
\KwOut{Set $P$ of unpacked paths $p$ with associated validity times $\Pi_p$}
\BlankLine

$P \gets \emptyset$\;
\For{$x \in X_{uv}$}{
  $[\tau_x, \pi_x] \gets T \cap \Pi_x$\;
  \eIf{$E(x_\tau) = uv \in A$}{
    $P \gets P \cup ([u,v], [\tau_x, \pi_x])$\;
  }{
    $w_x \gets E(x_\tau)$\;
    $P_{u w_x} \gets$ \unpack{$u w_x, [\tau_x, \pi_x]$}\;
    $P_{w_x v} \gets$ \unpack{$w_x v$, $[\tau_x + f_{u w_x}(\tau_x), \pi_x + f_{u w_x}(\pi_x)]$}\;
    $P_x \gets \combine{$P_{w_x v}, P_{u w_x}$}$
  }
  $P \gets P \cup P_x$
}

\Return{$P$}

\end{algorithm}

\section{Results}\label{sec:exp}

In this section, we present our experimental results.
We first discuss the experimental setup and the input road networks.
Then, we discuss the performance of each of our presented algorithms in turn.
Finally, we compare our approach to related work.

\subsection{Experimental Setup}

Our benchmark machine runs openSUSE Leap 15.2 (kernel 5.3.18), and has 192\,GiB of DDR4-2666 RAM and two Intel Xeon Gold 6144 CPUs, each of which has eight cores clocked at 3.5\,GHz and 8~$\times$~64\,KiB of L1, 8~$\times$~1\,MiB of L2, and 24.75\,MiB of shared L3 cache.
Hyperthreading was disabled and parallel experiments use 16 threads.
We implement our algorithms in Rust\footnote{The code for this paper and all experiments is available at \url{https://github.com/kit-algo/catchup}} and compile them with rustc 1.49.0-nightly (cf9cf7c92 2020-11-10) in the release profile with the target-cpu=native option\footnote{We disable AVX512 instructions, as they caused misoptimizations.}.
To compile competing implementations written in C++, we use GCC~9.3.1 using optimization level 3 and the -march=native option.

We investigated the performance of our preprocessing and query algorithms and compared it to competing algorithms.
Our experiments were focused on but not limited to space consumption and running times.
We performed preprocessing five times for each input network and report arithmetic means of the running times.
Unless reported otherwise, preprocessing utilized all 16 cores.
For queries, we generated 100\,000 source, target, departure time triples chosen uniformly at random for each graph.
These were executed in bulk.
Competing algorithms were evaluated with the same query set.
For profile queries, we only used 1000 queries (and discarded the departure time).
We report arithmetic means of query running times and machine independent measures such as number of nodes popped from the queue and number of evaluated travel time functions.

\subsection{Input Road Networks}

In this section, we report results for five road networks:
an instance of Berlin and the surrounding area, an old instance of Germany from 2006 which is used in many related works, two production-grade instances for Germany and Europe and with traffic predictions from 2017, and a recent production-grade instance of Europe.
All instances include traffic predictions as piecewise linear functions.
We use traffic predictions for a car on a typical midweek day or a Tuesday.
Table~\ref{tab:graphs} lists characteristics of each graph.

\begin{specialtable}
\centering
\caption{
Characteristics of test instances used.
The third column contains the percentage of arcs with a non-constant travel time function.
The fourth column the average number of breakpoints among those.
The fifth and sixth columns report the relative total delay for all/only non-constant arcs.
The final column contains the size of the graph representation in memory.
}\label{tab:graphs}
\begin{tabular}{lrrrrrrr}
\toprule
{} &         Nodes  &           Arcs & TD arcs & Avg. $|f|$ & Rel. Delay & Rel. Delay & Size \\
{} & [$\cdot 10^3$] & [$\cdot 10^3$] &    [\%] & per TD arc &       [\%] &    TD [\%] & [GB] \\
\midrule
Ber   &      443.2 &     988.5 &        27.4 &             75.0 &                   3.1 &                           17.6 &   0.2 \\
Ger06 &     4\,688.2 &   10\,795.8 &         7.2 &             19.5 &                   1.7 &                           33.1 &   0.3 \\
Ger17 &     7\,247.6 &   15\,752.1 &        29.2 &             31.6 &                   3.5 &                           20.8 &   1.3 \\
Eur17 &    25\,758.0 &   55\,503.8 &        27.2 &             29.5 &                   2.7 &                           19.0 &   4.2 \\
Eur20 &    28\,510.0 &   60\,898.8 &        76.3 &             22.5 &                  21.0 &                           34.9 &   8.7 \\
\bottomrule
\end{tabular}

\end{specialtable}

For time-independent routing, the performance of algorithms primarily depends on the size of the network.
In our case, however, the amount and complexity of time-dependent information also has a significant impact on the performance.
To measure this, we report the fraction of arcs which have a non-constant travel time and the average number of breakpoints among all non-constant travel time functions.
We also report the \emph{relative total delay} $\frac{\sum_a{\max{f_a} - \min{f_a}}}{\sum_a{\min{f_a}}}$ as a measure for the degree of time-dependency of the predictions\footnote{
Variants of this measure have been used in previous works.
Delling reported the average relative delay of time-dependent earliest arrival queries over the result of a time-independent query with lower bound travel times in~\cite{d-tdsr-11}.
Batz reported the average \emph{relative delay} $\frac{\max{f_a} - \min{f_a}}{\min{f_a}}$ in~\cite{bgsv-mtdtt-13}.
We use the \emph{total} delay because averages of ratios have hard to interpret semantics~\cite{hoefler2015scientific}.
For example, a short arc with a large relative delay could have a much bigger influence on the average relative delay than it has on the actual shortest path structure.
}.
The smaller the relative delay is, the greater is the effectiveness of pruning with upper and lower bounds.

The Ber instance was provided by TomTom\footnote{\url{https://www.tomtom.com}} in 2013.
It contains the northern eastern part of Germany, but in the literature it is referred to by the largest city, Berlin.
This instance features the most complex input functions and 27.4\% of the arcs have a non-constant travel time function.
However, the road network is quite small which makes the instance comparatively easy.
All other instances were provided by PTV\footnote{\url{https://ptvgroup.com}}.

The Ger06 instance is the easiest with respect to all time-dependency measures except the total relative delay limited to non-constant travel time functions.
That means that the degree of time-dependency in the non-constant functions is significant.
However, the overall influence on the shortest path structure is limited, because only 7.2\% of the arcs have a non-constant travel time.
Ger06 has also the smallest average complexity of non-constant travel time functions.

The newer graphs are not only bigger in terms of number of nodes and arcs but have also significantly more non-constant travel time functions.
Ger17 has four times as many time-dependent arcs as Ger06 and about 1.5 times as many breakpoints per time-dependent function.
Even though the relative delay among non-constant functions is not as high as for Ger06, the relative delay among all arcs is twice as high.
Eur17 exhibits similar characteristics but additionally has 3.5 times as many nodes and arcs.

Our newest instance is Eur20 with 28.5\,M nodes and 61\,M arcs.
Three quarters of the arcs have a non-constant travel time.
With around 35\%, the delay among non-constant functions is the greatest among all instances.
The total delay among all arcs is more than an order of magnitude higher than on Ger06.
This makes it the hardest instance.

We also performed experiments with predictions for different weekdays and the Western Europe graph provided for the 9th DIMACs implementation challenge~\cite{DemetrescuGJ09} with synthetic travel time functions~\cite{ndls-bastd-12}.
The results for these networks and prediction sets did not provide much additional insight.
We report them in Appendix~\ref{sec:exp_full} for the sake of comparability and completeness.

\subsection{Preprocessing}

\begin{specialtable}
\centering
\caption{Preprocessing statistics. Running times are for parallel execution with 16 cores.}\label{tab:prepro_data}
\begin{tabular}{lrrrrrrr}
\toprule
{} &       CCH arcs & \multicolumn{3}{c}{Expansions per arc} & Index & \multicolumn{2}{c}{Running time [s]} \\ \cmidrule(lr){3-5} \cmidrule(lr){7-8}
{} & [$\cdot 10^3$] &               Avg. & Max. & $= 1$ [\%] &  [GB] &                    Phase 1 & Phase 2 \\
\midrule
Ber   &         1\,977 &               1.039 &                  31 &           98.6 &       0.09 &                  1.5 &                  6.2 \\
Ger06 &        22\,519 &               1.075 &                  44 &           98.4 &       1.06 &                 30.1 &                 21.6 \\
Ger17 &        31\,488 &               1.090 &                 107 &           98.5 &       1.50 &                 35.0 &                107.4 \\
Eur17 &       114\,857 &               1.099 &                 115 &           98.4 &       5.47 &                189.6 &                557.0 \\
Eur20 &       128\,921 &               1.191 &                 109 &           96.9 &       6.32 &                209.6 &               1\,039.5 \\
\bottomrule
\end{tabular}

\end{specialtable}

Table~\ref{tab:prepro_data} reports the results for our preprocessing.
On Ger06, the first preprocessing step takes longer than the second.
However, for the newer instances with more time-dependent arcs and more breakpoints per function this changes and the second step becomes more expensive.
Despite that, the size of the final index corresponds only to the number of arcs in the augmented graph and does not grow as much for the newer instances.
The high complexity of the input function on Ber also does not depict any negative influence on the index size or the number of expansions per shortcut.
The augmented graphs have about twice as many arcs as the original graphs.
On average, only 1.1 expansions per arc need to be stored for all graphs (1.2 for Eur20).
About 98\% of all arcs have only a single expansion.
The maximum number of expansions per arc is only 115.
This is two orders of magnitude less than the number of breakpoints in the travel time function of that arc.
On Eur20, our hardest instance, the total preprocessing time is about 20 minutes, roughly twice as much as for Eur17.
However, the index size grows by less than 1\,GB and is in fact smaller than the input graph.
This clearly demonstrates the advantage in space efficiency of expansion information over explicitly storing travel time functions.

\subsubsection{Customization}\label{sec:exp:customization}

We now focus on the second preprocessing phase.
The first preprocessing phase does not use any time-dependent information and only reuses existing algorithms which have been evaluated in greater detail in other works~\cite{ghuw-fbndocch-19,z-cchtc-19}.
For the scope of this subsection, we only use the Eur20 graph.

To evaluate their impact, we selectively disable the triangle sorting and precustomization optimizations.
Even though both optimizations speed-up the customization by improving bounds, both have a significant impact on their own.
Disabling the precustomization increases the overall customization running time by about five minutes to 1311\,s.
The effect of triangle sorting is even stronger.
Disabling it roughly doubles customization running time to 2156\,s.

\subparagraph{Parallelization}

\begin{figure}
\centering
\includegraphics[width=\columnwidth]{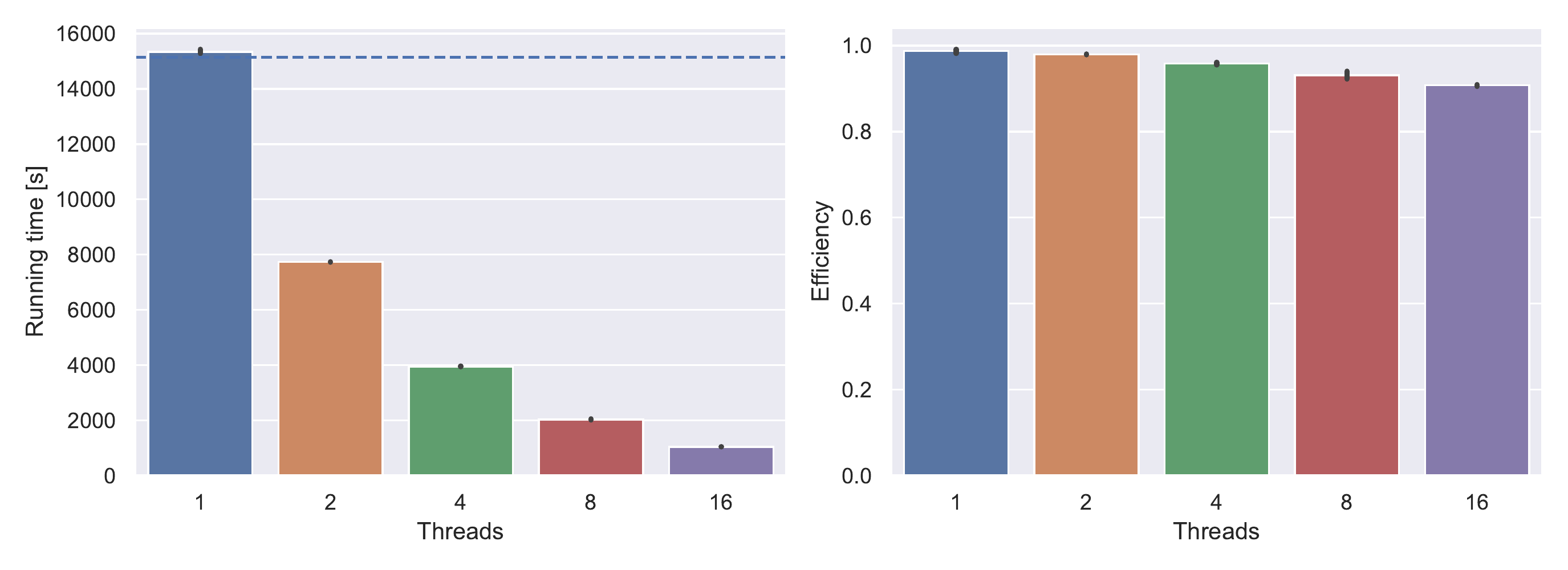}
\caption{
Average customization running times and parallelization efficiency (speedup/number of threads) on five customization runs of Eur20.
The black bars (barely visible) indicate the standard deviation.
The dashed line indicates running time with all parallelization code disabled.
}\label{fig:customization_scaling}
\end{figure}

We evaluate the effectiveness of our parallelization schema and run the customization with a varying number of threads.
Figure~\ref{fig:customization_scaling} depicts the results.
As a baseline, we run the experiment with all parallelization code disabled.
The baseline running time is indicated by the dashed line.
Enabling parallelization but running with only one thread causes only little overhead.
Running with more threads introduces more overhead due to synchronization.
With 16 threads, parallel efficiency is still around 0.9.
We conclude that our parallelization schema scales well and that customization times could be reduced further by utilizing additional cores.

\begin{figure}
\widefigure
\centering
\includegraphics[width=.85\columnwidth]{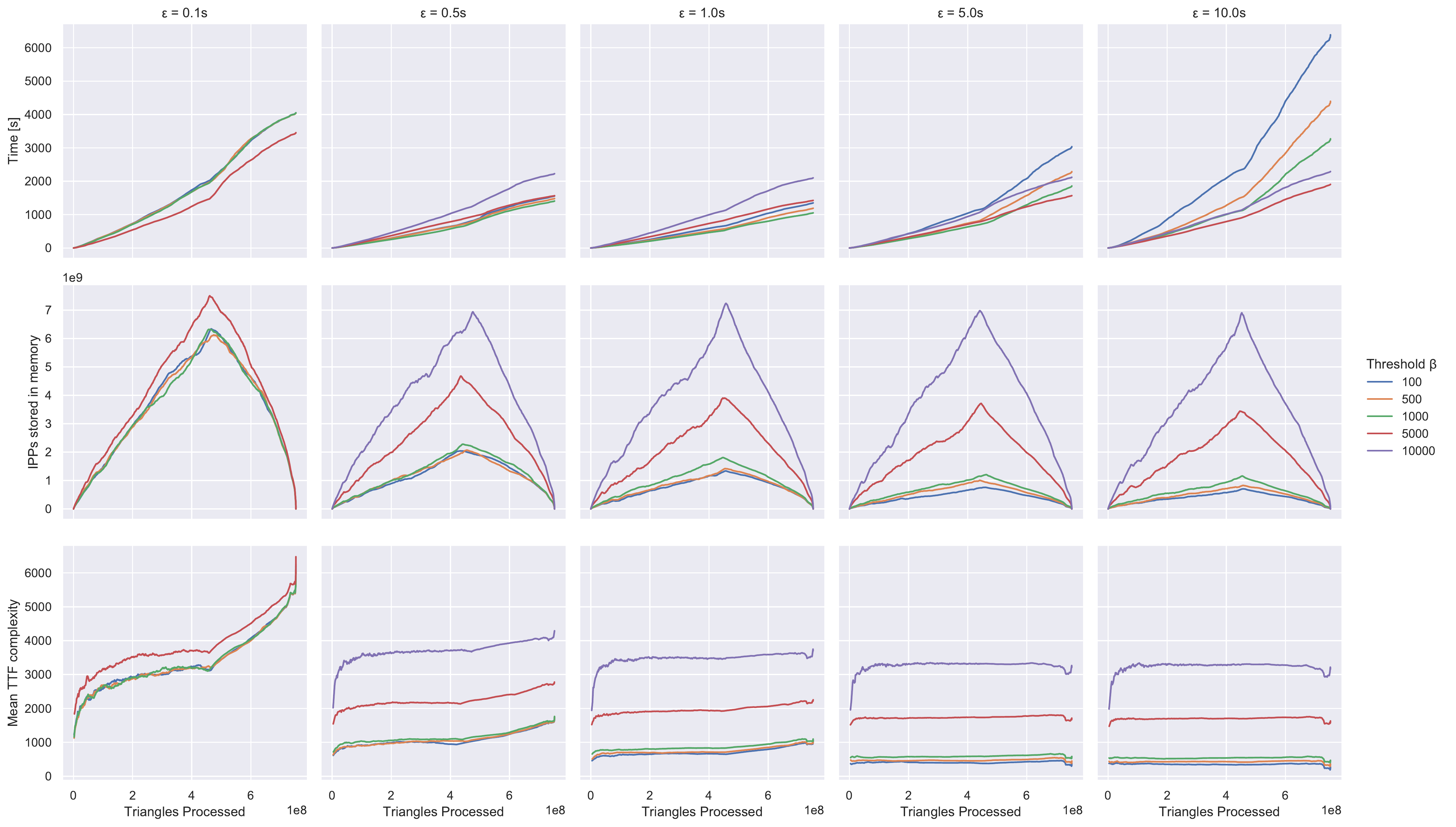}
\caption{
Customization behavior depending on the approximation parameters difference $\epsilon$ (varying by column) and threshold $\beta$ (indicated by color).
The x-axis in all plots indicates the progress of the customization by number of processed triangles.
The y-axis is the passed time in the first row, the memory usage in the second row (measure by the total number of stored breakpoints), and the mean travel time function complexity in the third row.
There are many more elements which contribute to memory consumption.
However, the breakpoints for travel time functions are the biggest chunk and are the easiest to measure.
A breakpoint has a size of 16 bytes in memory.
Thus, $8 \cdot 10^9$ breakpoints correspond to 128\,GB memory consumption for travel time functions alone.
The configuration $\beta = 10000$, $\epsilon = 0.1s$ caused an out-of-memory error and is not listed.
}\label{fig:customization_stats}
\end{figure}

\subparagraph{Approximation}
We perform customization experiments with different approximation parameters.
Over the course of the customization, we track the progress over time, the memory consumption and the average travel time function complexity.
Figure~\ref{fig:customization_stats} displays these measurements.
We use the number of processed triangles to measure the progress because it corresponds roughly linearly to the time passed (though different parameters lead to different slopes).
After about 60\% of the triangles, the slope changes slightly.
At this point only high-level separator nodes/arcs remain.
These have complex travel time functions so linking and merging becomes more expensive.
Also, we switch from task based to loop based parallelization which is less effective.
Measuring progress by processed nodes (the for loop in Line~1 in Algorithm~\ref{alg:customization}) or processed arcs (for loop in Line~2) is also insightful.
However, for these, the correspondence is not linear.
The last couple of thousand nodes and the last million arcs take almost half the total time.

We observe that the choice of approximation parameters has a huge influence on running time and memory consumption.
The best running time of around 1000\,s is achieved with $\beta = 1000$ and $\epsilon = 1.0s$.
Thus, we use it as our default configuration.
The worst running time is over six times higher.
In the best configuration, we use only around 25\,GB for travel time functions while a bad parameter choice or no approximation leads to crashes with out-of-memory errors.
Generally, a larger $\epsilon$ leads to looser approximation, lower travel time function complexity, and thus less memory consumption.
Conversely, a larger $\beta$ causes the approximation to be executed less often and the memory consumption and function complexity increases.
The average function complexity is usually well below $\beta$ except for very small values of $\epsilon$ or $\beta$.
In that case the complexity cannot be reduced sufficiently to keep the complexity below $\beta$.
If $\beta$ is large, approximation is performed seldom and the influence of $\epsilon$ becomes smaller.
Similarly, when $\epsilon$ is small, the influence of $\beta$ is limited because the complexity cannot be reduced enough no matter how often approximation is performed.
Clearly, the right choice of approximation parameters is essential to the performance of the preprocessing.
When travel time functions are too complex, too much memory is used, and linking and merging are very expensive.
However, when functions are approximated too loosely, a lot of time is spent in the reconstruction of exact functions for the times when bounds overlap.
Thus, extreme parameter choice for $\epsilon$ and $\beta$ are detrimental to the running time, even though they may reduce memory consumption.

Through all configurations, the memory usage peaks after around 60\% of the triangles have been processed.
The reason is the way we maintain travel time functions in memory during the customization.
An arc's travel time function is stored once its lower ranked endpoint has been processed until its higher ranked endpoint has been processed.
In the beginning, we process nodes with low rank and store many travel time functions.
Only once we reach the higher ranked nodes, we start dropping a significant amount of the stored functions.
This causes the observed peak.

\startlandscape

\finishlandscape

\subsection{Queries}

\begin{specialtable}
\centering
\caption{
Query performance with different optimizations.
We report the number of nodes popped from the queue, the number of evaluated travel time functions (TTFs) and the running time.
All values are arithmetic means over 100\,000 queries executed in bulk with source, target and departure time drawn uniformly at random.
}\label{tab:queries}
\begin{tabular}{llrrr}
\toprule
{} & {} & Queue & Evaluated &   Running \\
{} & {} &  pops &      TTFs & time [ms] \\
\midrule
Ber & Basic &              167.4 &                   100\,820.5 &              8.8 \\
      & + Corridor &               38.1 &                     5\,224.1 &              0.6 \\
      & + Lazy &             1\,603.6 &                     1\,747.4 &              0.6 \\
      & + A* &              635.2 &                      691.5 &              0.3 \\
\addlinespace
Ger06 & Basic &              492.3 &                   818\,721.3 &             46.4 \\
      & + Corridor &               79.7 &                    31\,740.8 &              2.3 \\
      & + Lazy &             3\,323.2 &                     3\,838.0 &              1.7 \\
      & + A* &              831.0 &                      995.1 &              0.6 \\
\addlinespace
Ger17 & Basic &              510.3 &                  2\,100\,731.8 &            169.7 \\
      & + Corridor &              143.4 &                   164\,372.5 &             13.7 \\
      & + Lazy &            18\,450.0 &                    19\,910.5 &              9.1 \\
      & + A* &             3\,099.2 &                     3\,495.5 &              1.7 \\
\addlinespace
Eur17 & Basic &              861.6 &                  9\,951\,623.1 &            808.6 \\
      & + Corridor &              229.3 &                   806\,727.8 &             62.3 \\
      & + Lazy &            39\,714.8 &                    43\,581.1 &             20.8 \\
      & + A* &             6\,876.5 &                     7\,911.0 &              4.1 \\
\addlinespace
Eur20 & Basic &              871.0 &                 10\,527\,072.7 &            813.2 \\
      & + Corridor &              335.6 &                  1\,222\,655.6 &             92.9 \\
      & + Lazy &            62\,677.7 &                    70\,145.4 &             33.7 \\
      & + A* &             7\,231.9 &                     8\,844.7 &              4.7 \\
\bottomrule
\end{tabular}

\end{specialtable}

In this section, we investigate the performance of our query algorithms.
Table~\ref{tab:queries} depicts the influence of the query optimizations.
The basic approach does not achieve competitive running times.
Queries take almost a second on average on the newer Europe graphs.
Naively evaluating shortcut travel times with Algorithm~\ref{alg:eval} is too slow.

Limiting the search space to a shortest path corridor using the elimination tree interval query significantly reduces running times.
The speedup is between a factor of 20 on Ger06 and 8 and Eur20.
The effectiveness of this optimization corresponds inversely to the relative delay (see Table~\ref{tab:graphs}).
Greater relative delays lead to bigger corridors and thus smaller speedups.

The lazy evaluation optimization has limited impact on the running time (speedups between 1.3 and 3).
However, it drastically shifts the balance between queue operations and travel time function evaluations.
On the newer graphs, the number of queue pops increases by more than two orders of magnitude, while the number of travel time function evaluations decreases by up to a factor of 20.
The additional queue operations introduce some overhead.
However, this is mitigated by the avoided unnecessary and duplicate evaluations.
Reusing the lower bounds from the corridor search for goal directed search yields an additional speedup of factor two to seven.

\subsubsection{Local Queries}

\begin{figure}
\centering
\includegraphics[width=\columnwidth]{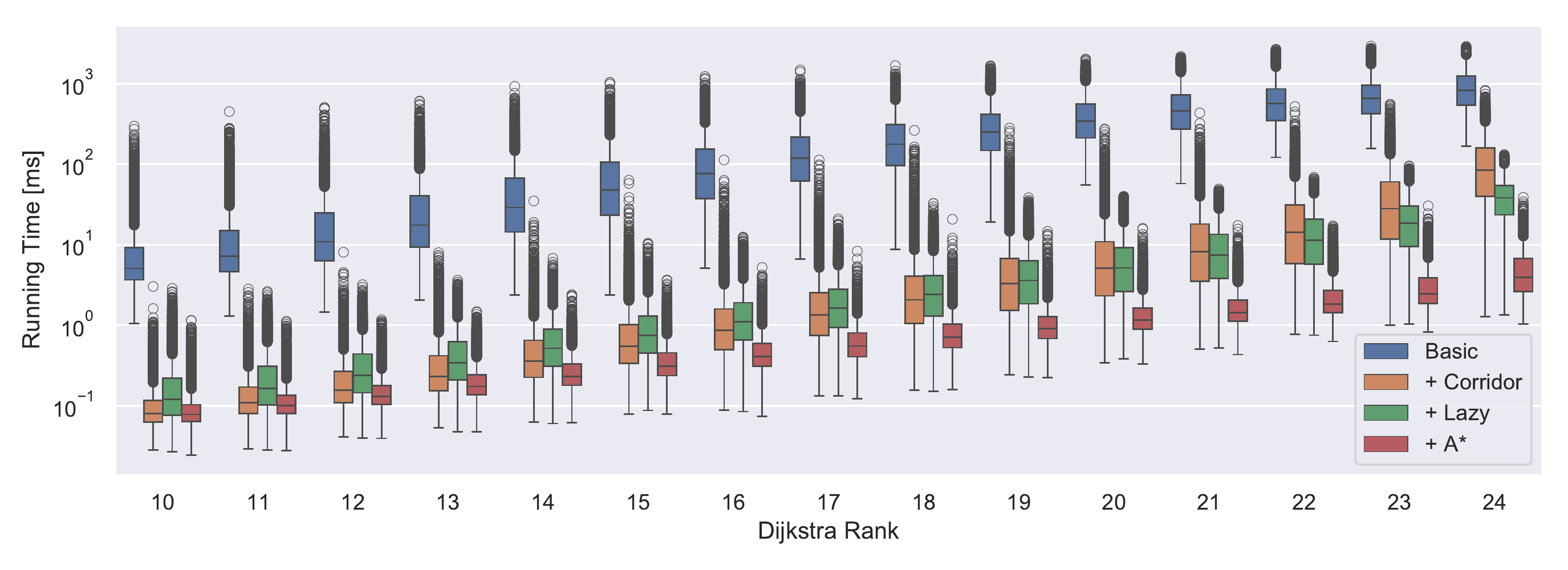}
\caption{Query running times in milliseconds with different optimizations by Dijkstra ranks on Eur20. The boxes cover the range between the first and the third quartile. The band in the box indicates the median. The whiskers indicate 1.5 times the inter quartile range. Running times outside this range are considered as outliers and displayed separately.}\label{fig:ranks}
\end{figure}

We generate another set of queries to investigate the performance of our algorithms depending on the distance of source and target.
We draw 10\,000 start nodes and departure times uniformly at random and perform time-dependent Dijkstra without a specific target.
For every $2^i$th settled node, we store it as the target of a query of \emph{Dijkstra rank} $i$.
This methodology was introduced by \citet{ss-hhhes-05}.
Figure~\ref{fig:ranks} shows query running times by rank for the query algorithm with the various query optimizations enabled successively.

Obviously, query running times scale with the distance.
The fully optimized algorithm takes only fractions of milliseconds for short range queries, except for some outliers which take up to a millisecond.
For long range queries, we usually achieve query times within a couple of milliseconds and the maximum query time was 39\,ms.
The basic query algorithm is around two orders of magnitude slower across all ranks.
The impact of lazy optimization appears to depend on the rank of the query.
For lower ranks, it introduces some overhead but reduces outliers compared to only the corridor optimization.
This is due to the overhead of the queue operations.
For long range queries, this is completely amortized by the reduction in arc relaxations.

\subsubsection{Profile Queries}\label{sec:exp:profile_queries}

\begin{specialtable}
\centering
\caption{
Running times of profile queries and characteristics of the obtained profiles.
We report total running times and running times of each phase (Corridor, Reconstruction, Contraction, Extraction) of the query.
The total running time is slightly larger than the sum of all phases as it includes some additional initialization and cleanup work.
We report the number of breakpoints in the obtained travel time profile (Column $|f|$).
Column $|X|$ contains the number of times the shortest path changes during the day.
Since the same path may be the fastest for several times, we also report the number of distinct paths in the last column.
All values are arithmetic means over 1000 queries executed in bulk with source and target nodes drawn uniformly at random.
}\label{tab:profile_queries}
\begin{tabular}{lrrrrrrrrr}
\toprule
{} & \multicolumn{6}{c}{Running time [ms]}         &    {} &    {} &       {} \\ \cmidrule(lr){2-7}
{} & I & II & III & \multicolumn{2}{c}{IV} & Total & $|f|$ & $|X|$ & Distinct \\ \cmidrule(lr){5-6}
{} &   &    &     &            TTF & Paths &       &       &       &    paths \\
\midrule
Ber   &                    0.1 &              37.8 &           13.9 &                 2.8 &                0.4 &             55.6 &             30\,974.9 &            2.7 &                 2.3 \\
Ger06 &                    0.4 &              56.5 &           23.0 &                 0.7 &                0.8 &             83.6 &              9\,359.6 &            6.9 &                 3.3 \\
Ger17 &                    0.8 &             452.2 &          189.0 &                 6.1 &                2.4 &            660.1 &             66\,146.0 &            9.7 &                 3.7 \\
Eur17 &                    1.7 &            1\,135.8 &          732.9 &                13.0 &                7.8 &           1\,913.2 &            122\,192.1 &           16.5 &                 6.8 \\
Eur20 &                    2.8 &            3\,166.6 &         1\,507.7 &                12.3 &               10.3 &           4\,747.5 &            107\,690.7 &           24.4 &                11.6 \\
\bottomrule
\end{tabular}

\end{specialtable}

We perform experiments for profile queries and report the results in Table~\ref{tab:profile_queries}.
The total running time depends on the amount of time-dependent information in the instance.
From Ger06 to Ger17 the total running time increase by a factor of 8 even though the network grows only little.
The same can be observed between Eur17 and Eur20.

The total running time is dominated by the reconstruction and contraction phases.
Reconstruction of the travel time functions of the existing shortcuts takes roughly twice as long as computing the functions for the new shortcuts in the contraction phase.
The corridor phase with the elimination tree interval query takes a negligible amount of time.
Surprisingly, the time required to compute a final exact travel time profile is not much greater than to compute a path profile (on Ger06, the path profile is even slower than the travel time profile).

The average complexity of the final travel time function varies by an order of magnitude across the different instances.
This confirms that Ger06 is a relatively simple instance.
Surprisingly, the average complexity on Eur17 is higher than on Eur20.
We suspect that this is caused by the higher average complexity in the input graph and that many important arcs that cover many shortest paths already have a non-constant travel time function in Eur17.
The number of path switches is magnitudes smaller.
It ranges from three path switches on Ber to 24 on Eur20.
The number of distinct paths during the day are roughly half of that.
The arithmetic mean of these numbers is slightly skewed upwards by a few very high values.
However, the median is still fairly close:
For example, the median number of distinct paths is two on Ger06 and ten on Eur20.

\subsection{Comparison with Related Work}

Table~\ref{tab:related_work} provides an overview over different techniques, their preprocessing and query times, space overhead of the index data structures and average query errors where approximation is used.
Where possible, we obtained the code of competing algorithms\footnote{KaTCH: \url{https://github.com/GVeitBatz/KaTCH}\\ TD-S: \url{https://github.com/ben-strasser/td_p}} and evaluated them with same methodology, instances and queries as our algorithms.
For other competitors, we report available numbers from the respective publications.

\begin{specialtable}[t!]
\centering
\caption{
Comparison with related work.
We list unscaled numbers as reported in the respective publications for algorithms we could not run ourselves.
Values not reported are indicated as n/r. 
OOM means that the program crashed while trying to allocate more memory than available.
A similar overview with scaled numbers can be found in \protect\cite{d-earpm-16}.
}\label{tab:related_work}
\begin{tabular}{ll@{\hskip10pt}rrrrrrrr}
\toprule{} & {} & \multicolumn{2}{c}{Preprocessing} & Index & \multicolumn{3}{c}{Query} \\ \cmidrule(lr){3-4} \cmidrule(lr){6-8}
{} & {} & Time & Cores & size &    Time & \multicolumn{2}{c}{Rel. error} \\ \cmidrule(lr){7-8}{} & {} &  [s] &       & [GB] &    [ms] & Avg. [\%] & Max. [\%] \\
\midrule
\parbox[t]{3mm}{\multirow{18}{*}{\rotatebox[origin=c]{90}{Ger06}}} & TD-Dijkstra &                              - &     - &               - &               719.26 &                    - &                    - \\
      & TDCALT \cite{dn-crdtd-12} &                            540 &     1 &   \textbf{0.23} &                 5.36 &                    - &                    - \\
      & TDCALT-K1.15 \cite{dn-crdtd-12} &                            540 &     1 &   \textbf{0.23} &                 1.87 &                0.050 &               13.840 \\
      & eco L-SHARC \cite{d-tdsr-11} &                           4\,680 &     1 &            1.03 &                 6.31 &                    - &                    - \\
      & heu SHARC \cite{d-tdsr-11} &                          12\,360 &     1 &            0.64 &                 0.69 &                  n/r &                0.610 \\
      & KaTCH &                            169 &    16 &            4.66 &                 0.64 &                    - &                    - \\
      & TCH \cite{bgsv-mtdtt-13} &                            378 &     8 &            4.66 &                 0.75 &                    - &                    - \\
      & ATCH (1.0) \cite{bgsv-mtdtt-13} &                            378 &     8 &            1.12 &                 1.24 &                    - &                    - \\
      & ATCH ($\infty$) \cite{bgsv-mtdtt-13} &                            378 &     8 &            0.55 &                 1.66 &                    - &                    - \\
      & inex. TCH (0.1) \cite{bgsv-mtdtt-13} &                            378 &     8 &            1.34 &                 0.70 &                0.020 &                0.100 \\
      & inex. TCH (1.0) \cite{bgsv-mtdtt-13} &                            378 &     8 &            1.00 &                 0.69 &                0.270 &                1.010 \\
      & TD-CRP (0.1) \cite{bdpw-dtdrp-16} &                            289 &    16 &            0.78 &                 1.92 &                0.050 &                0.250 \\
      & TD-CRP (1.0) \cite{bdpw-dtdrp-16} &                            281 &    16 &            0.36 &                 1.66 &                0.680 &                2.850 \\
      & FLAT \cite{kppwz-iotdr-17a} &                         158\,760 &     6 &           54.63 &                 1.27 &                0.015 &                  n/r \\
      & CFLAT \cite{kppwz-iotdr-17a} &                         104\,220 &     6 &           34.63 &        \textbf{0.58} &                0.008 &                0.918 \\
      & TD-S+9 &                            542 &     1 &            3.61 &                 2.07 &                0.001 &                1.523 \\
      & \textbf{\tdcch{}} &                    \textbf{52} &    16 &            1.06 &                 0.72 &                    - &                    - \\
\addlinespace \parbox[t]{3mm}{\multirow{4}{*}{\rotatebox[origin=c]{90}{Ger17}}} & TD-Dijkstra &                              - &     - &               - &               814.60 &                    - &                    - \\
      & KaTCH &                            859 &    16 &           42.81 &        \textbf{1.26} &                    - &                    - \\
      & TD-S+9 &                            601 &     1 &            5.28 &                 2.61 &                0.001 &                0.963 \\
      & \textbf{\tdcch{}} &                   \textbf{142} &    16 &   \textbf{1.50} &                 2.02 &                    - &                    - \\
\addlinespace \parbox[t]{3mm}{\multirow{4}{*}{\rotatebox[origin=c]{90}{Eur17}}} & TD-Dijkstra &                              - &     - &               - &              2\,929.72 &                    - &                    - \\
      & KaTCH &                           3\,066 &    16 &          146.97 &                  OOM &                    - &                    - \\
      & TD-S+9 &                           3\,149 &     1 &           18.84 &        \textbf{4.70} &                0.002 &                1.159 \\
      & \textbf{\tdcch{}} &                   \textbf{747} &    16 &   \textbf{5.47} &                 4.92 &                    - &                    - \\
\addlinespace \parbox[t]{3mm}{\multirow{4}{*}{\rotatebox[origin=c]{90}{Eur20}}} & TD-Dijkstra &                              - &     - &               - &              3\,784.11 &                    - &                    - \\
      & KaTCH &                           7\,149 &    16 &          239.78 &                  OOM &                    - &                    - \\
      & TD-S+9 &                           3\,352 &     1 &           20.65 &        \textbf{4.23} &                0.006 &                1.733 \\
      & \textbf{\tdcch{}} &                  \textbf{1\,249} &    16 &   \textbf{6.32} &                 5.60 &                    - &                    - \\
\bottomrule
\end{tabular}

\end{specialtable}

In our comparison, KaTCH, heu SHARC, CFLAT and \tdcch{} all achieve query times around 0.6\,ms on Ger06.
The original research implementation TCH reports slightly slower times than KaTCH.
This may be because experiments were run on an older machine, but also because according to the KaTCH documentation, the newer query is somewhat more efficient.
TCH pays for this speed with 4.7\,GB index data.
Reducing the KaTCH memory consumption while keeping exactness (ATCH) brings query times up to 1.24\,ms.
ATCH also feature a configuration where they only keep upper and lower bounds for each travel time function (ATCH $\infty$).
This configuration uses even less memory than \tdcch{} because the optimized order results in fewer shortcuts.
However, query running times degrade to 1.66\,ms.
Giving up on exactness allows keeping the query times at 0.7\,ms (inex. TCH) but introduces noticeable errors.

While achieving competitive query times for acceptable memory consumption, heu SHARC suffers from huge preprocessing times of several hours.
The original publication does not report average query errors, only a maximum error of 0.61\%.
TDCALT has the smallest memory consumption but does not achieve competitive query times, even when approximating.
FLAT and CFLAT both suffer from extreme preprocessing times and memory consumption despite having no exact queries.
\tdcch{} offers competitive query times for exact results while keeping memory consumption reasonable.
TD-CRP offers even lower memory consumption. 
However, this is only possible through the use of approximation.
TD-CRP queries depict a noticeable error and perform somewhat worse than KaTCH or \tdcch{} queries.
TD-S+9 depicts the smallest average error of all non-exact approaches\footnote{\citet{kppwz-iotdr-17a} reported another CFLAT configuration with even smaller errors but significantly slower queries.}.

Path retrieval in the time-dependent scenario is not as easy as in the static setting.
Table~\ref{tab:related_work} reports running times for the earliest arrival query and the path retrieval combined.
We only have separate numbers for KaTCH and \tdcch{}. 
For CFLAT, \citet{kppwz-iotdr-17a} reported that path retrieval takes a third of the total query time.
Our experiments show a similar amount for KaTCH.
For \tdcch{}, path retrieval takes up less than 10\% of the query time.
TD-CRP and FLAT do not support path retrieval.

On Ger17, KaTCH query times increase by a factor of about two.
However, memory usage grows by almost an order of magnitude.
For TD-S, both the growth in space consumption and query times corresponds roughly to the growth of the graph size, but not to the increased number of breakpoints.
The index of \tdcch{} grows by a similar factor.
Query times get about 2.7 times slower.

On Eur17, the memory consumption of KaTCH becomes prohibitive.
While KaTCH is still able to finish preprocessing and output 150\,GB of data, queries crash since the 192\,GB RAM of our machine are not enough.
Using ATCH or inexact TCH, the memory consumption could likely be reduced sufficiently to perform queries.
However, this would either introduce errors or slow down queries significantly.
With only 5.5\,GB of index data, \tdcch{} is still able to perform exact queries in less than 5\,ms on average.
This is fast enough to enable interactive applications.
Total preprocessing for \tdcch{} takes less than a quarter of the time KaTCH needs.
TD-S+9 is also able to handle this instance with similar query times but only with a small average error.

On Eur20, the behavior is similar, only more extreme.
KaTCH preprocessing time increases by more than a factor of two and index data grows to 240\,GB.
The TD-S+9 numbers remain relatively stable.
Query times get slightly faster but errors become larger.
\tdcch{} preprocessing times also become slower but by less than a factor of two.
Query times increase to 5.6\,ms.
The index takes only 6.3\,GB, which is smaller than the input network.

\section{Discussion}\label{sec:conclusion}

We introduce \tdcch{}, a speed-up technique for routing in time-dependent road networks.
It features a small index size and fast, exact queries.
To the best of our knowledge, our approach is the first to simultaneously achieve all three objectives.
We perform an extensive experimental study to evaluate the performance of \tdcch{} and compare it to competing approaches.
Our approach achieves the fastest preprocessing, competitive query running times and up to 38 times smaller indexes than other approaches.
This clearly demonstrates the advantage of storing expansion information instead of travel time functions.

Revisiting ATCH, TCH, and TD-CRP with the insights gained in this work could be fruitful.
Combining ATCH with our A* query extension could reduce ATCH query running times.
\tdcch{} makes use of travel time independent node orders.
Combining \tdcch{} with TCH-like node orders could result in even smaller index sizes and query running times.
We further expect that some of our optimizations to the preprocessing can also be applied in a TD-CRP context.
Another possible direction for future research would be to support partial updates to further accelerate the customization.
This could enable the integration of live traffic information.

\vspace{6pt}

\authorcontributions{Conceptualization, B.S. and T.Z.; methodology, B.S. and T.Z.; software, T.Z.; validation, B.S. and T.Z.; formal analysis, T.Z.; investigation, B.S. and T.Z.; resources, D.W.; data curation, B.S. and T.Z.; writing---original draft preparation, B.S. and T.Z.; writing---review and editing, B.S., D.W., and T.Z.; visualization, B.S. and T.Z.; supervision, D.W.; project administration, D.W.; funding acquisition, D.W. All authors have read and agreed to the published version of the manuscript.}

\funding{
This research was funded by Karlsruhe Institute of Technology and BMW Group.
}

\acknowledgments{
We thank Lars Gottesb\"uren and Michael Hamann for fruitful discussions and feedback.
We also thank Marcel Radermacher for his input on approximation algorithms.
}

\appendixtitles{yes} 
\appendix

\renewcommand{\thespecialtable}{A\arabic{specialtable}}

\section{Extended Experimental Results}\label{sec:exp_full}

In this section, we document our experimental results on additional networks and prediction sets for different weekdays.
Table~\ref{tab:graphs_full} contains characteristics for the full set.
We report results for predictions for different weekdays on Ber and Ger06.
Judging from the numbers in Table~\ref{tab:graphs_full}, the additional weekdays are no harder than the midweek predictions.
The SynEur instance uses the Western Europe graph provided for the 9th DIMACs implementation challenge~\cite{DemetrescuGJ09} with synthetic travel time functions~\cite{ndls-bastd-12}.
The evaluation in~\cite{bgsv-mtdtt-13} also used the medium and high traffic prediction sets.
In comparison to the real world data sets, SynEur features extremely high delays on its time-dependent arcs.
However, there are only few time-dependent arcs.
Even the high traffic set has less time-dependent arcs than Ger06 midweek.
This combination of few time-dependent arcs with extremely high travel time fluctuations causes some interesting effects in our experiments.
Since the behavior of CATCHUp is very stable across all real world data sets, we are confident that this is an artifact of the synthetic predictions.

Table~\ref{tab:prepro_full} contains preprocessing results for all graphs.
The other days for Ger06 and Ber behave roughly as expected.
The weekend instances feature less time-dependent arcs and preprocessing accordingly runs faster.
SynEur with medium and high traffic produces some surprising results regarding the unpacking data.
Even the medium traffic instance has a higher average number of expansions than Eur17.
With high traffic, the number is even greater than on Eur20.
The number of arcs with only a single expansion is correspondingly small.
We suspect that the reason for this are the extreme relative delays of the predicted travel time functions.
These extreme fluctuations in travel time lead to many shortest path changes despite the little amount of time-dependent information.

The results for query experiments on all instances reported in Table~\ref{tab:queries_full} also confirm our observations from the main part of this article.
Each optimization yields similar accelerations.
Again, SynEur exhibits surprising behavior.
On the one hand, unoptimized queries are surprisingly fast, i.e., up to four times faster than on Eur17.
On the other hand, with all optimizations, SynEur with high traffic has the slowest query times among all instances.
Again, the reason is the combination of few time-dependent arcs with high relative delays.
Because there is little time-dependent information in the instance, the basic query algorithm is not as slow as one could expect.
However, because of the high delays, the corridor search and the A* optimizations are not as effective.
In~\cite{ndls-bastd-12} it is stated that unimportant arcs (with respect to a Highway Hierarchy) will never get a non-constant travel time function.
In combination with the high relative delays, detours through unimportant parts of the network can often become the shortest paths.
This also decreases the effectiveness of our A* optimization.

Table~\ref{tab:profile_queries_full} contains profile query results for all instances.
Once again, the results mostly conform to the already reported observations and SynEur deviates.
In this case, the results are particularly surprising.
While the travel time profiles are comparatively simple because of the low complexity of the input functions, the number of path switches is so high that we initially suspected bugs as the cause.
In addition, it decreases as the amount of traffic increases.
Nevertheless, we claim that the numbers are correct and that the reason lies in the combination of high relative delays with few time-dependent arcs.
When there are only few time-dependent arcs and the slowdown due to a predicted traffic jam on an arc is very high, there will always be a faster detour using less important arcs without travel time predictions.
This leads to the extremely high number of switches and distinct paths.
As the amount of time-dependent arcs is increased, the spatial consistency increases and an increasing amount of detours will now also have an increased travel time.
Thus, the number of path switches decreases.
\end{paracol}

\appendix

\begin{specialtable}
\widetable
\centering
\caption{
Characteristics of test instances used.
The third column contains the percentage of arcs with a non-constant travel time function.
The fourth column the average number of breakpoints among those.
The fifth and sixth columns report the relative total delay for all/non-constant arcs.
The final column contains the size of the graph representation in memory.
}\label{tab:graphs_full}
\begin{tabular}{llrrrrrrr}
\toprule
{} & {} &         Nodes  &           Arcs & TD arcs & Avg. $|f|$ & Rel. Delay & Rel. Delay & Size \\
{} & {} & [$\cdot 10^3$] & [$\cdot 10^3$] &    [\%] & per TD arc &       [\%] &    TD [\%] & [GB] \\
\midrule
Ber & Monday &      443.2 &     988.5 &        27.4 &             74.6 &                   3.1 &                           17.7 &   0.2 \\
      & Tuesday &      443.2 &     988.5 &        27.4 &             75.0 &                   3.1 &                           17.6 &   0.2 \\
      & Wednesday &      443.2 &     988.5 &        27.5 &             74.9 &                   3.1 &                           17.5 &   0.2 \\
      & Thursday &      443.2 &     988.5 &        27.6 &             75.2 &                   3.2 &                           17.7 &   0.2 \\
      & Friday &      443.2 &     988.5 &        27.2 &             73.4 &                   3.1 &                           17.5 &   0.2 \\
      & Saturday &      443.2 &     988.5 &        20.2 &             69.1 &                   2.1 &                           14.8 &   0.1 \\
      & Sunday &      443.2 &     988.5 &        19.9 &             67.2 &                   2.0 &                           14.6 &   0.1 \\
\addlinespace
Ger06 & Monday &     4\,688.2 &   10\,795.8 &         7.0 &             20.1 &                   1.7 &                           33.1 &   0.3 \\
      & midweek &     4\,688.2 &   10\,795.8 &         7.2 &             19.5 &                   1.7 &                           33.1 &   0.3 \\
      & Friday &     4\,688.2 &   10\,795.8 &         6.4 &             18.9 &                   1.5 &                           32.0 &   0.3 \\
      & Saturday &     4\,688.2 &   10\,795.8 &         3.9 &             15.8 &                   0.8 &                           28.5 &   0.2 \\
      & Sunday &     4\,688.2 &   10\,795.8 &         2.5 &             15.0 &                   0.4 &                           26.2 &   0.2 \\
\addlinespace
SynEur & Low &    18\,010.2 &   42\,188.7 &         0.1 &             13.2 &                   0.3 &                          125.2 &   0.8 \\
      & Medium &    18\,010.2 &   42\,188.7 &         1.0 &             13.2 &                   0.8 &                          124.9 &   0.8 \\
      & High &    18\,010.2 &   42\,188.7 &         6.2 &             13.2 &                   4.6 &                          124.8 &   1.0 \\
\addlinespace
Ger17 & Tuesday &     7\,247.6 &   15\,752.1 &        29.2 &             31.6 &                   3.5 &                           20.8 &   1.3 \\
\addlinespace
Eur17 & Tuesday &    25\,758.0 &   55\,503.8 &        27.2 &             29.5 &                   2.7 &                           19.0 &   4.2 \\
\addlinespace
Eur20 & Tuesday &    28\,510.0 &   60\,898.8 &        76.3 &             22.5 &                  21.0 &                           34.9 &   8.7 \\
\bottomrule
\end{tabular}

\end{specialtable}

\begin{specialtable}
\widetable
\centering
\caption{Preprocessing statistics. Running times are for parallel execution on 16 cores.}\label{tab:prepro_full}
\begin{tabular}{llrrrrrrr}
\toprule
{} & {} &       CCH arcs & \multicolumn{3}{c}{Expansions per arc} & Index & \multicolumn{2}{c}{Running time [s]} \\ \cmidrule(lr){4-6} \cmidrule(lr){8-9}
{} & {} & [$\cdot 10^3$] &               Avg. & Max. & $= 1$ [\%] &  [GB] &                    Phase 1 & Phase 2 \\
\midrule
Ber & Monday &         1\,977 &               1.040 &                  25 &           98.6 &       0.09 &                  1.5 &                  6.2 \\
      & Tuesday &         1\,977 &               1.039 &                  31 &           98.6 &       0.09 &                  1.5 &                  6.2 \\
      & Wednesday &         1\,976 &               1.038 &                  19 &           98.6 &       0.09 &                  1.5 &                  6.2 \\
      & Thursday &         1\,977 &               1.039 &                  23 &           98.6 &       0.09 &                  1.6 &                  6.2 \\
      & Friday &         1\,975 &               1.037 &                  28 &           98.7 &       0.09 &                  1.5 &                  5.8 \\
      & Saturday &         1\,961 &               1.023 &                  21 &           99.1 &       0.09 &                  1.5 &                  3.8 \\
      & Sunday &         1\,957 &               1.021 &                  27 &           99.2 &       0.09 &                  1.6 &                  3.3 \\
\addlinespace
Ger06 & Monday &        22\,499 &               1.073 &                  42 &           98.4 &       1.06 &                 30.0 &                 20.9 \\
      & midweek &        22\,519 &               1.075 &                  44 &           98.4 &       1.06 &                 30.1 &                 21.6 \\
      & Friday &        22\,445 &               1.064 &                  43 &           98.6 &       1.05 &                 30.2 &                 17.2 \\
      & Saturday &        22\,229 &               1.031 &                  37 &           99.2 &       1.03 &                 30.2 &                  6.0 \\
      & Sunday &        22\,128 &               1.019 &                  39 &           99.5 &       1.02 &                 29.8 &                  3.6 \\
\addlinespace
SynEur & Low &        88\,884 &               1.036 &                  23 &           99.2 &       4.14 &                238.3 &                 82.7 \\
      & Medium &        90\,514 &               1.109 &                  24 &           97.6 &       4.31 &                231.5 &                224.8 \\
      & High &        94\,302 &               1.264 &                  31 &           94.6 &       4.71 &                233.3 &                523.0 \\
\addlinespace
Ger17 & Tuesday &        31\,488 &               1.090 &                 107 &           98.5 &       1.50 &                 35.0 &                107.4 \\
\addlinespace
Eur17 & Tuesday &       114\,857 &               1.099 &                 115 &           98.4 &       5.47 &                189.6 &                557.0 \\
\addlinespace
Eur20 & Tuesday &       128\,921 &               1.191 &                 109 &           96.9 &       6.32 &                209.6 &               1\,039.5 \\
\bottomrule
\end{tabular}

\end{specialtable}
\begin{paracol}{2}
\switchcolumn
\startlandscape
\begin{specialtable}
\centering
\caption{
Query performance with different optimizations.
We report the number of nodes popped from the queue, the number of evaluated travel time functions and the running time.
All values are arithmetic means over 100\,000 queries executed in bulk with source, target and departure time drawn uniformly at random.
}\label{tab:queries_full}
\begin{tabular}{llrrrrrrrrrrrr}
\toprule
{} & {} & \multicolumn{4}{c}{Queue pops} & \multicolumn{4}{c}{Evaluated travel time functions} & \multicolumn{4}{c}{Running time [ms]} \\ \cmidrule(lr){3-6} \cmidrule(lr){7-10} \cmidrule(lr){11-14}
{} & {} & Basic & + Corridor & + Lazy & + A* & Basic & + Corridor & + Lazy & + A* & Basic & + Corridor & + Lazy & + A* \\
\midrule
Ber & Monday &             167.4 &       38.2 &  1\,605.0 &  618.6 &                   99\,629.0 &     5\,480.3 &  1\,762.4 &   674.9 &             8.6 &        0.6 &    0.6 &  0.3 \\
      & Tuesday &             167.4 &       38.1 &  1\,603.6 &  635.2 &                  100\,820.5 &     5\,224.1 &  1\,747.4 &   691.5 &             8.8 &        0.6 &    0.6 &  0.3 \\
      & Wednesday &             167.4 &       38.6 &  1\,640.4 &  643.7 &                  101\,938.1 &     5\,405.3 &  1\,786.6 &   702.0 &             8.9 &        0.6 &    0.7 &  0.3 \\
      & Thursday &             167.4 &       38.9 &  1\,647.9 &  642.5 &                  101\,584.1 &     5\,498.9 &  1\,799.8 &   701.8 &             8.8 &        0.6 &    0.7 &  0.3 \\
      & Friday &             167.4 &       37.8 &  1\,591.1 &  619.6 &                   99\,142.5 &     5\,061.0 &  1\,722.9 &   674.0 &             8.5 &        0.5 &    0.6 &  0.3 \\
      & Saturday &             167.1 &       26.1 &   926.5 &  491.5 &                   86\,470.8 &     2\,124.5 &   967.4 &   514.1 &             6.4 &        0.3 &    0.3 &  0.2 \\
      & Sunday &             167.1 &       24.7 &   864.1 &  476.3 &                   84\,796.6 &     1\,865.9 &   895.3 &   495.1 &             6.1 &        0.2 &    0.3 &  0.2 \\
\addlinespace
Ger06 & Monday &             492.3 &       68.6 &  2\,649.0 &  727.0 &                  751\,679.3 &    22\,542.0 &  3\,029.0 &   853.0 &            42.8 &        1.8 &    1.4 &  0.5 \\
      & midweek &             492.3 &       79.7 &  3\,323.2 &  831.0 &                  818\,721.3 &    31\,740.8 &  3\,838.0 &   995.1 &            46.4 &        2.3 &    1.7 &  0.6 \\
      & Friday &             491.9 &       62.9 &  2\,349.2 &  731.3 &                  780\,031.8 &    21\,423.2 &  2\,665.4 &   848.3 &            42.6 &        1.6 &    1.2 &  0.5 \\
      & Saturday &             490.7 &       24.1 &   339.4 &  211.1 &                  541\,331.4 &     2\,457.5 &   360.0 &   223.6 &            26.8 &        0.4 &    0.3 &  0.2 \\
      & Sunday &             490.0 &       20.2 &   219.2 &  163.5 &                  503\,009.4 &     1\,599.4 &   226.8 &   169.1 &            24.2 &        0.3 &    0.2 &  0.2 \\
\addlinespace
SynEur & Low &             742.8 &      341.3 &  6\,626.2 & 1\,704.8 &                 4\,871\,967.5 &   997\,409.9 & 16\,521.1 &  4\,730.7 &           201.8 &       39.9 &    5.1 &  2.0 \\
      & Medium &             746.8 &      461.8 & 17\,209.3 & 3\,796.9 &                 5\,742\,442.6 &  1\,596\,401.3 & 35\,066.3 &  9\,389.8 &           253.0 &       69.4 &   13.1 &  4.0 \\
      & High &             749.7 &      554.1 & 33\,572.2 & 7\,018.4 &                 6\,142\,257.3 &  2\,031\,399.6 & 60\,234.2 & 15\,685.2 &           289.2 &       96.4 &   25.5 &  6.9 \\
\addlinespace
Ger17 & Tuesday &             510.3 &      143.4 & 18\,450.0 & 3\,099.2 &                 2\,100\,731.8 &   164\,372.5 & 19\,910.5 &  3\,495.5 &           169.7 &       13.7 &    9.1 &  1.7 \\
\addlinespace
Eur17 & Tuesday &             861.6 &      229.3 & 39\,714.8 & 6\,876.5 &                 9\,951\,623.1 &   806\,727.8 & 43\,581.1 &  7\,911.0 &           808.6 &       62.3 &   20.8 &  4.1 \\
\addlinespace
Eur20 & Tuesday &             871.0 &      335.6 & 62\,677.7 & 7\,231.9 &                10\,527\,072.7 &  1\,222\,655.6 & 70\,145.4 &  8\,844.7 &           813.2 &       92.9 &   33.7 &  4.7 \\
\bottomrule
\end{tabular}

\end{specialtable}

\begin{specialtable}
\centering
\caption{
Running times of profile queries and characteristics of the obtained profiles.
We report total running times and running times of each phase (Corridor, Reconstruction, Contraction, Extraction) of the query.
The total running time is slightly larger than the sum of all phases as it includes some additional initialization and cleanup work.
We report the number of breakpoints in the obtained travel time profile (Column $|f|$).
Column $|X|$ contains the number of times the shortest path changes during the day.
Since the same path may be the fastest for several times, we also report the number of distinct paths in the last column.
All values are arithmetic means over 1000 queries executed in bulk with source and target nodes drawn uniformly at random.
}\label{tab:profile_queries_full}
\begin{tabular}{llrrrrrrrrr}
\toprule
{} & {} & \multicolumn{6}{c}{Running time [ms]}                       &    {} &    {} &       {} \\ \cmidrule(lr){3-8}
{} & {} & Corridor & Reconstruct & Contract &   Exact & Paths & Total & $|f|$ & $|X|$ & Distinct \\
{} & {} &          &             &          & profile &       &       &       &       &    paths \\
\midrule
Ber & Monday &                    0.1 &              39.7 &           14.9 &                 2.6 &                0.4 &             58.4 &             29\,090.5 &            2.7 &                 2.3 \\
      & Tuesday &                    0.1 &              37.8 &           13.9 &                 2.8 &                0.4 &             55.6 &             30\,974.9 &            2.7 &                 2.3 \\
      & Wednesday &                    0.1 &              38.4 &           14.1 &                 2.8 &                0.4 &             56.5 &             31\,126.2 &            2.6 &                 2.2 \\
      & Thursday &                    0.1 &              40.1 &           14.8 &                 2.8 &                0.4 &             58.8 &             30\,662.1 &            2.7 &                 2.3 \\
      & Friday &                    0.1 &              34.2 &           12.5 &                 2.4 &                0.3 &             50.2 &             27\,671.5 &            2.6 &                 2.2 \\
      & Saturday &                    0.1 &              11.2 &            3.4 &                 1.5 &                0.2 &             16.6 &             17\,892.8 &            1.6 &                 1.8 \\
      & Sunday &                    0.1 &               8.9 &            2.5 &                 1.4 &                0.2 &             13.2 &             16\,768.7 &            1.6 &                 1.8 \\
\addlinespace
Ger06 & Monday &                    0.3 &              42.2 &           17.3 &                 0.6 &                0.8 &             62.9 &              9\,036.7 &            7.0 &                 3.1 \\
      & midweek &                    0.4 &              56.5 &           23.0 &                 0.7 &                0.8 &             83.6 &              9\,359.6 &            6.9 &                 3.3 \\
      & Friday &                    0.3 &              32.4 &           14.0 &                 0.5 &                0.7 &             49.3 &              7\,896.2 &            6.0 &                 3.0 \\
      & Saturday &                    0.2 &               1.7 &            0.9 &                 0.1 &                0.4 &              3.6 &              2\,047.2 &            2.8 &                 2.0 \\
      & Sunday &                    0.2 &               0.9 &            0.4 &                 0.1 &                0.3 &              2.1 &              1\,386.3 &            2.2 &                 1.8 \\
\addlinespace
SynEur & Low &                    2.1 &             591.3 &          501.7 &                 0.2 &                7.3 &           1\,128.2 &              3\,379.4 &          151.4 &               144.9 \\
      & Medium &                    3.1 &            1\,664.3 &          962.7 &                 0.4 &                6.1 &           2\,694.1 &              5\,226.1 &           99.0 &                90.1 \\
      & High &                    3.2 &            3\,521.5 &         1\,484.9 &                 0.5 &                5.3 &           5\,102.2 &              5\,939.1 &           78.0 &                69.2 \\
\addlinespace
Ger17 & Tuesday &                    0.8 &             452.2 &          189.0 &                 6.1 &                2.4 &            660.1 &             66\,146.0 &            9.7 &                 3.7 \\
\addlinespace
Eur17 & Tuesday &                    1.7 &            1\,135.8 &          732.9 &                13.0 &                7.8 &           1\,913.2 &            122\,192.1 &           16.5 &                 6.8 \\
\addlinespace
Eur20 & Tuesday &                    2.8 &            3\,166.6 &         1\,507.7 &                12.3 &               10.3 &           4\,747.5 &            107\,690.7 &           24.4 &                11.6 \\
\bottomrule
\end{tabular}

\end{specialtable}
\finishlandscape
\end{paracol}
\reftitle{References}
\externalbibliography{yes}

\end{document}